\documentclass[twoside,english]{iopart}
\usepackage[T1]{fontenc}
\usepackage[latin9]{inputenc}
\usepackage[a4paper]{geometry}
\usepackage{color}
\usepackage{babel}
\usepackage{float}
\usepackage{bm}
\usepackage{graphicx}
\usepackage{esint}
\usepackage[unicode=true,pdfusetitle,
 bookmarks=true,bookmarksnumbered=false,bookmarksopen=false,
 breaklinks=false,pdfborder={0 0 1},backref=false,colorlinks=true]
 {hyperref}
\usepackage{breakurl}

\makeatletter

\providecommand{\tabularnewline}{\\}

\usepackage{iopams}
\usepackage{setstack}

\usepackage[numbers, sort&compress]{natbib}

\newcommand{\eqref}[1]{(\ref{#1})}

\makeatother

\begin{document}

\title{Surface effects on ferromagnetic resonance in magnetic nanocubes}

\author{{\normalsize{}R. Bastardis$^{1}$, F. Vernay$^{1}$, D.-A. Garanin$^{2}$
and H. Kachkachi$^{1}$}}

\address{$^{1}$Laboratoire PROMES CNRS (UPR-8521), Université de Perpignan
Via Domitia, Rambla de la thermodynamique, Tecnosud, F-66100 Perpignan,
France}

\address{$^{2}$Physics Department, Lehman College, City University of New
York 250 Bedford Park Boulevard West, Bronx, New York 10468-1589,
USA}

\ead{roland.bastardis@univ-perp.fr}
\begin{abstract}
We study the effect of surface anisotropy on the spectrum of spin-wave
excitations in a magnetic nanocluster and compute the corresponding
absorbed power. For this, we develop a general numerical method based
on the (undamped) Landau-Lifshitz equation, either linearized around
the equilibrium state leading to an eigenvalue problem or solved using
a symplectic technique. For box-shaped clusters, the numerical results
are favorably compared to those of the finite-size linear spin-wave
theory. Our numerical method allows us to disentangle the contributions
of the core and surface spins to the spectral weight and absorbed
power. In regard to the recent developments in synthesis and characterization
of assemblies of well defined nano-elements, we study the effects
of free boundaries and surface anisotropy on the spin-wave spectrum
in iron nanocubes and give orders of magnitude of the expected spin-wave
resonances. For an $8\ {\rm nm}$ iron nanocube, we show that the
absorbed power spectrum should exhibit a low-energy peak around 10
GHz, typical of the uniform mode, followed by other low-energy features
that couple to the uniform mode but with a stronger contribution from
the surface. There are also high-frequency exchange-mode peaks around
60 GHz.
\end{abstract}
\maketitle

\section{Introduction}

During the last decades the development of potential technological
applications of magnetic nanoparticles, such as magnetic imaging and
magnetic hyperthermia, has triggered a new endeavor for a better control
of the relevant properties of such systems. In particular, synthesis
and growth of crystalline nanoparticles have reached such a high level
of skill and know-how as to produce well defined $2D$ and $3D$ arrays
of nanoclusters of tailored size, shape and internal crystal structure
\cite{sun00science,LisieckiAdv_Mat,tartajetal03jpd,LisieckiJPhCS,Respaud_nanocube2008,Respaud_nanocube2_2010JMMM}.
On the other hand, experimental measurements on nanoscale systems
are a step behind inasmuch as they still do not provide us with sufficient
space-time resolutions for an unambiguous interpretation of the observed
phenomena that are commonly attributed to finite-size or surface effects.
Nonetheless, ferromagnetic resonance (FMR), which is a well known
and very precise technique for characterizing bulk and layered magnetic
media \cite{vonsovskii66pp,gurmel96crcpress,heinrich94springer},
benefits from a renewed interest in the context of nanomagnetism.
Indeed, some newly devised variants of the FMR technique \cite{tran06phd,leeetal11jap,kronastetal11nl,goncalvesetal13apl,Schoeppneretal14jap,ollefsetal15jap}
combine the study of dynamic magnetic properties by FMR with the elemental
specificity of the chemical composition of the particles. For instance,
these techniques can be employed to detect the ferromagnetic resonance
of single Fe nanocubes with a sensitivity of $10^{6}\mu_{\mathrm{B}}$
and element-specific excitations in Co-Permalloy structures. Another
variant of ferromagnetic resonance spectroscopy is the so-called Magnetic
Resonance Force Microscopy (MRFM)\cite{Sidles_RevModPhys}. It has
recently been used for the characterization of cobalt nanospheres
\cite{deLoubens_ArXiv2014}. These techniques hold the prospect of
providing a better resolution of the surface properties at the level
of a single (isolated) magnetic nanoparticle. For the benefits of
theoretical work, these experiments could provide the missing data
for resolving the surface response to a time-dependent magnetic field,
and thus contribute to assess the validity of surface-anisotropy models.
In particular, measurements of the absorbed power in FMR experiments
on ``isolated'' particles or dilute assemblies of nanoparticles
could serve these purposes. Indeed, this is a standard observable
that is routinely measured in such experiments. From the theory standpoint,
it is a well known (dynamic) response of a magnetic system that can
be computed by various well established techniques, analytical as
well as numerical.

In the present work we consider a box-shaped nanocluster modeled as
a many-spin system with free boundary conditions, subjected to a time-dependent
(small-amplitude) magnetic field. The systems considered here are
chosen to model, to some extent, Fe nanocubes studied by several groups
\cite{tartajetal03jpd,tran06phd,trunovaetal08jap,Respaud_nanocube2008,Respaud_nanocube2_2010JMMM}.
Our main objective is to distinguish and assess the role of surface
and core contributions to the FMR absorption spectra. For this we
focus on the simple system of an isolated (ferromagnetic) nanocube
and study its intrinsic properties, thus ignoring its interactions
with other nanocubes that would be included in an assembly and its
interactions with the hosting matrix. As shown by Sukhov \textit{et
al}. \cite{sukhovetal08j3m}, this assumption is fully justified in
the case of dilute samples. Obviously, real systems of magnetic nanoparticles
are far more complex. Indeed, Fe nanoparticles may present a variety
of morphologies and internal structures, especially in a core/shell
configuration where one observes an antiferromagnetic layer coating
a ferromagnetic system \cite{briaticoetal99epjd,lingetal09nanolett,lacroixetal11nanolett}.
However, the system we adopt is simple enough to illustrate our study
in a clear manner but rich enough to capture the main physics we are
interested in. Furthermore, the methods we develop here are quite
versatile and can be extended to a given magnetic nanoparticle with
arbitrary physical parameters.

Consequently, the energy of the nanocluster considered here includes
the Zeeman energy, the (nearest-neighbor) spin-spin exchange coupling
and on-site anisotropy (core and surface). We also allow for the possibility
that exchange interactions involving one or more sites in the surface
outer shell to be different from those in the core or at the interface
between the core and the surface. Upon solving the (undamped) Landau-Lifshitz
equation (LLE) we compute the absorbed power of such systems. Then,
the LLE equation is linearized around the equilibrium state of lowest
energy and the ensuing eigenvalue problem is solved to infer the full
spectrum (eigenfrequencies and eigenfunctions) of all spin-wave excitations.
Finally, by comparison with the absorbed power of a given mode, we
can determine the separate contributions of core and surface of the
nanocluster.

The paper is organized as follows : in the next Section we present
our model and computing methods. We give the model Hamiltonian and
then describe the two numerical methods we used to compute the full
spin-wave spectrum (eigenfrequencies and eigenvectors) and the absorbed
power. In Section III we present and discuss our results for the effects
of size and surface anisotropy on the absorbed power. This section
ends with a discussion of Fe nanocubes for which we give orders of
magnitude and speculate on the possibility to observe the calculated
peak in the absorbed power. An appendix has been added on a toy model
of a three-layer system in order to illustrate, in a simpler manner,
how the various branches in the spin-wave dispersion can be associated
with spins of a given type (core or surface) in the system.

\section{Model and Methods}

\subsection{The Hamiltonian }

We model the magnetic nanocluster as a system of $\mathcal{N}$ classical
spins $\mathbf{s}_{i}$, with $|\mathbf{s}_{i}|=1$, with the help
of the Hamiltonian 
\begin{equation}
\mathcal{H}=\mathcal{H}_{{\rm ex}}+\mathcal{H}_{{\rm an}},\label{eq:ham_general}
\end{equation}
where 
\begin{equation}
\mathcal{H}_{{\rm ex}}=-\frac{1}{2}\sum_{i,j}J_{ij}\mathbf{s}_{i}\cdot\mathbf{s}_{j}\label{eq:ham_ex}
\end{equation}
is the ferromagnetic Heisenberg exchange interaction and $\mathcal{H}_{{\rm an}}$
the anisotropy contribution. We assume only nearest-neighbor interactions
(nn), $J_{ij}=J>0$ for $i,j\in\mathrm{nn}$ and zero otherwise. However,
we use a numerical method that allows us to consider different exchange
couplings between the core and surface spins according to their loci.
More precisely, we may distinguish between core-core ($J_{c}$), core-surface
($J_{cs}$) and surface-surface ($J_{s}$) exchange coupling. The
anisotropy term in Eq. (\ref{eq:ham_general}) is assumed to be uniaxial
(along the $z$ axis) with constant $D>0$ for core spins and of Néel's
type with constant $D_{S}$ for surface spins. More precisely, the
anisotropy energy is local (on-site), so that $\mathcal{H}_{{\rm an}}=\sum_{i}\mathcal{H}_{{\rm an},i}$,
and given by 
\begin{equation}
\mathcal{H}_{{\rm an},i}=\left\{ \begin{array}{lc}
-D\left(\mathbf{s}_{i}\cdot\mathbf{e}_{z}\right)^{2}, & i\in\mathrm{core}\\
\\
\frac{1}{2}D_{S}{\displaystyle \sum_{j\in\mathrm{nn}}}\left(\mathbf{s}_{i}\cdot\mathbf{u}_{ij}\right)^{2}, & i\in\mathrm{surface}.
\end{array}\right.\label{eq:ham_an}
\end{equation}

Here $\mathbf{u}_{ij}$ is the unit vector connecting the surface
site $i$ to its nearest neighbor on site $j$. Néel's anisotropy
arises due to missing nearest neighbors for surface spins. In particular,
for the simple cubic lattice and $xy$ surfaces (perpendicular to
the $z$ axis), the Néel anisotropy becomes $\mathcal{H}_{{\rm an},i}=-\frac{1}{2}D_{S}s_{i,z}^{2}$.
This means that for $D_{S}>0$ the spins tend to align perpendicularly
to the surface, while for $D_{S}<0$ the surface spins tend to align
in the tangent plane. In a box-shaped nanocluster the Néel anisotropy
on the edges along the $z$ axis becomes $\mathcal{H}_{{\rm an},i}=-\frac{1}{2}D_{S}\left(s_{i,x}^{2}+s_{i,y}^{2}\right)$
or, equivalently, $\mathcal{H}_{{\rm an},i}=\frac{1}{2}D_{S}s_{i,z}^{2}$.
As such, for $D_{S}>0$ the edge spins tend to align perpendicularly
to the edges. On the other hand, it is easy to check that Néel's anisotropy
vanishes at the corners and in the core of a box-shaped nanocluster.

For the sake of simplicity, and for an easier comparison with experiments
on iron nanocubes, for instance, the systems investigated in the present
work are boxed-shaped with $\mathcal{N}=N_{x}\times N_{y}\times N_{z}$
with a simple cubic lattice. In this case, the surface anisotropy
(SA) favours an ordering along the shortest edges of the particle
if $D_{S}>0$ and along the longest ones otherwise. Indeed, for an
atom on the edge in the $x$ direction, for instance, we have $4$
neighbors with $\mathbf{u}_{ij}=\mathbf{e}_{x},\mathbf{u}_{ij}=-\mathbf{e}_{x},\mathbf{u}_{ij}=\mathbf{e}_{y},\mathbf{u}_{ij}=-\mathbf{e}_{z}$
and thereby (using $|\mathbf{s}_{i}|=1$) we obtain $\mathcal{H}_{{\rm an},i}\rightarrow\frac{D_{S}}{2}+\frac{D_{S}}{2}\, s_{i,x}^{2}$.

\subsection{Excitation spectrum and absorbed power: computing methods}

\label{sec:Dynamic}

Since surface anisotropy is much stronger than the core anisotropy
and the fraction of surface spins for nanoclusters is appreciable,
SA strongly influences the spin-wave spectrum of the cluster. Experimentally,
the most accessible modes are the spin-wave modes that couple to the
uniform ac field, as in magnetic-resonance experiments. In the absence
of SA, only the uniform-precession mode is seen in the magnetic resonance.
The effect of SA is twofold. First, the uniform (or nearly uniform)
precession frequency is modified by SA; it increases or decreases
depending on geometry. Hence, combining magnetic-resonance experiments
with the corresponding theoretical results provides a means for estimating
the surface-anisotropy constant. Second, in larger clusters exchange
stiffness becomes less restrictive and different groups of spins (such
as the core and surface spins) can precess at different frequencies
and this leads to several resonance peaks.

In this Section we describe two complementary methods we have used
to compute the spin-wave spectrum and the absorbed power. The first
method consists in linearizing the (undamped) Landau-Lifshitz equation
(LLE) around the equilibrium position and then solving the ensuing
eigenvalue problem to obtain the eigenfrequencies and the corresponding
eigenvectors (spin-wave modes). This method is quite versatile as
it can be applied to any nanocluster with arbitrary size, shape and
energy parameters. In the case of box-shaped nanoclusters this method
is compared with the results of linear spin-wave theory obtained in
Refs. \cite{kacgar01physa300, kacgar01epjb}. The second numerical
method used here consists in directly solving the LLE using the technique
of symplectic integrators \cite{krechetal98cpc,stesch06cpc}. As will
be seen later these two methods are in a very good agreement.

\subsubsection{Linearization of the Landau-Lifshitz equation: normal modes of a
nanocluster\label{sub:LinLLE}}

Here we deal with the numerical solution of the Landau-Lifshitz equation
(LLE) of motion 
\begin{equation}
\hbar\frac{d{\bf s}_{i}}{dt}={\bf s}_{i}\times\mathbf{H}_{\mathrm{eff},i}-\lambda\,{\bf s}_{i}\times\left({\bf s}_{i}\times\mathbf{H}_{\mathrm{eff},i}\right),\label{eq:LL_eq}
\end{equation}
where the effective field is defined by $\mathbf{H}_{\mathrm{eff},i}=-\delta_{\mathbf{s}_{i}}\mathcal{H}+g\mu_{B}\mathbf{H}(t)$,
with $g$ being the Landé factor and $\mu_{B}$ the Bohr magneton,
$\lambda$ the dimensionless damping parameter and $\mathbf{H}(t)$
the time-dependent magnetic field. In the following we set $\lambda=0$
(Larmor equation) to avoid artificial effects. Internal spin-wave
processes in the particle can provide a natural damping of spin waves,
especially for larger particles and non-zero temperatures. For nanosize
particles, spin-wave modes are essentially discrete \cite{kacgar01physa300,kacgar01epjb},
while damping requires quasi-continuous excitation branches to satisfy
energy conservation in spin-wave processes. In addition, we do not
include thermal excitation via stochastic Langevin fields in the model.
Thus we expect that the spin wave modes of our particles are undamped.
In other words, in this work, we are not seeking the precise result
for the microwave absorption. We use these calculations to find positions
of spin-wave peaks and compare them with a second approach. Our numerical
experiment is short-time whereas damping comes into play at longer
times that we are not considering here.

One of the goals of the present work is to assess the role of the
surface contribution to the energy spectrum of a single nanocluster
or to a given physical observable that is easily accessible experimentally,
\emph{e.g. }the absorbed power. So, before we compute the relevant
observable, it is necessary to compute the eigenvectors and eigenenergies
of the system. Then, it is our aim to try to attribute the various
peaks in the energy spectrum to the core or surface contributions
and to estimate the corresponding statistical weight. The eigenvalue
problem by linearizing the LLE (\ref{eq:LL_eq}) around the equilibrium
state $\left\{ {\bf s}_{i}^{(0)}\right\} _{i=1,\ldots,\mathcal{N}}$.
This has been done in the system of spherical coordinates in order
to reduce the number of equations from $3\mathcal{N}$ to $2\mathcal{N}$.
The main steps of our formalism are summarized in \ref{sec:Spherical-coordinates}.
More precisely, we write $\delta{\bf s}_{i}={\bf s}_{i}-{\bf s}_{i}^{(0)}$,
for $i=1,\ldots,\mathcal{N}$, and expand the first derivative of
the energy $\mathcal{E}$ (or the effective field) to $1^{\mathrm{st}}$-order
in $\delta{\bf s}_{i}$ 
\begin{equation}
\begin{array}{lll}
\mathbf{H}_{\mathrm{eff},i}\left\{ {\bf s}^{(0)}+\delta{\bf s}\right\}  & = & \mathbf{H}_{\mathrm{eff},i}\left\{ {\bf s}^{(0)}\right\} +\left[\sum\limits _{j=1}^{\mathcal{N}}\left(\delta{\bf s}_{j}\mathbf{\cdot\bm{\nabla}}_{j}\right)\mathbf{H}_{\mathrm{eff},i}\right]\left\{ {\bf s}^{(0)}\right\} .\end{array}\label{eq:ExpandedEffField}
\end{equation}

Then, inserting this into the LLE (\ref{eq:LL_eq}) leads to 
\begin{equation}
\frac{d\left(\delta{\bf s}_{i}\right)}{dt}=\sum\limits _{j=1}^{\mathcal{N}}\left[\bm{\widetilde{\mathcal{H}}}_{ij}\mathcal{I}\right]\delta{\bf s}_{j},\quad i=1,\ldots,\mathcal{N}\label{eq:LinearizedLLEMatrixForm}
\end{equation}
where $\bm{\widetilde{\mathcal{H}}}_{ik}$ is the pseudo-Hessian defined
in Eq. (\ref{eq:SC-pseudoHessian}) and 
\[
\mathcal{I}\equiv\left(\begin{array}{cc}
0 & -1\\
1 & 0
\end{array}\right)
\]
\textcolor{black}{is a matrix} that results from the vector product
of ${\bf s}_{i}$ with the effective field $\mathbf{H}_{\mathrm{eff},i}$.
The solution of Eq. (\ref{eq:LinearizedLLEMatrixForm}) can be sought
in the form $\delta{\bf s}_{k}\left(t\right)=\delta{\bf s}_{k}\left(0\right)\exp\left(i\omega t\right)$,
leading to the eigenvalue problem 
\begin{eqnarray}
\sum\limits _{j=1}^{\mathcal{N}}\left(\bm{\widetilde{\mathcal{H}}}_{ij}\mathcal{I}-i\omega\mathbf{1}\right)\delta{\bf s}_{j} & = & 0,\label{eq:eigenvalue_pb}
\end{eqnarray}
whose solution yields the excitation spectrum of the nanocluster.
Accordingly, the eigenvalue problem (\ref{eq:eigenvalue_pb}) is then
solved numerically for an arbitrary $\mathcal{N}$-spin nanocluster
by diagonalizing the $2\mathcal{N}\times2\mathcal{N}$ matrix with
elements $\left[\bm{\widetilde{\mathcal{H}}}_{ij}\mathcal{I}\right]^{\alpha\beta}$.
This is done in the absence of the time-dependent magnetic field $\mathbf{H}(t)$
so that the effective field involved here is given by $\mathbf{H}_{\mathrm{eff},i}=-\delta_{\mathbf{s}_{i}}\mathcal{H}$.

In order to evaluate the contributions of the surface and core spins
to the eigenvector (or mode) $\delta\mathbf{s}_{k}$, we introduce
the corresponding ``spectral weight''. For this purpose, we first
write the eigenvector $\delta\mathbf{s}_{k}$ of wave vector $\bm{k}$
as

\begin{equation}
\delta{\bf s}_{k}\left(0\right)=\sum_{i=1}^{\mathcal{N}}f_{ki}\delta{\bf s}_{i}\left(0\right)\label{eq:FourierCoeffs}
\end{equation}
with $f_{ki}$ are the eigenfunctions of the matrix $\left[\bm{\widetilde{\mathcal{H}}}\mathcal{I}\right]$.
For later use the equation above can be rewritten as 
\begin{equation}
\delta\mathbf{s}_{k}\left(0\right)=\sum_{i=1}^{\mathcal{N}}\sum_{\alpha=x,y,z}\mathcal{D}_{ki}^{\alpha}{\bf e}_{i}^{\alpha}\label{eq:coeff_spins}
\end{equation}
where $\left\{ {\bf e}_{i,x},{\bf e}_{i,y},{\bf e}_{i,z}\right\} $
is the local Cartesian frame and $\mathcal{D}_{ki}^{\alpha}$ are
the corresponding coefficients.

Then, we may define the spectral weight (per site) associated with
the core and surface spins as follows

\[
W_{k}^{s,c}=\frac{1}{\mathcal{N}}\times\left(\frac{1}{N_{s,c}}\sum_{i\in\mathrm{core,}\mathrm{surface}}|f_{ki}|^{2}\right)
\]
with the normalization condition $N_{s}W_{k}^{s}+N_{c}W_{k}^{c}=1$,
where $N_{c}\left(N_{s}\right)$ is the number of core (surface) spins. 

In Ref. \cite{kacgar01epjb} the eigenfunctions $f_{ki}$ were calculated
analytically using the finite-size spin-wave theory for a boxed-shaped
particle. This yields a benchmark for the numerical results obtained
here and helps interpret them. The spin-wave excitations were treated
perturbatively as small deviations of the spins $\mathbf{s}_{i}$
from the direction ${\bf n}$ of the particle's net magnetic moment,
namely $\mathbf{s}_{i}\simeq{\bf n}+\mathbf{\bm{\pi}}_{i}$, with
${\bf n}\cdot\mathbf{\bm{\pi}}_{i}=0$. The eigenfunctions were then
obtained in the form

\begin{equation}
\mathbf{\pi}_{k}=\sum_{ix,iy,iz}\left(f_{ix,kx}\times f_{iy,ky}\times f_{iz,kz}\right)\mathbf{\pi}_{i}\label{eq:Spinwave_function}
\end{equation}
with

\begin{equation}
f_{i\alpha,k\alpha}=\sqrt{\frac{2}{1+\delta_{k\alpha}}}\cos\left[\left(i_{\alpha}-1/2\right)k_{\alpha}\right],\quad\alpha=x,y,z
\end{equation}
and $k_{\alpha}=\frac{n_{\alpha}\pi}{N_{\alpha}}$, in the case of
free boundary conditions, as adopted here. Comparing Eq. (\ref{eq:Spinwave_function})
with Eq. (\ref{eq:FourierCoeffs}) we see that the variables $\mathbf{\bm{\pi}}_{k}$
used in Ref. \cite{kacgar01epjb} are in fact identical to the variables
$\delta\mathbf{s}_{k}$ defined in Eq. (\ref{eq:FourierCoeffs}).

The normal modes of a magnetic nanocluster have been studied by many
authors {[}see Refs. \cite{grimsditchetal04prb69, grimsditchetal04prb70}
and references therein{]}. On the other hand, the ferromagnetic resonance
of ensembles of magnetic nanoparticles in the macrospin approximation
has also been studied numerically using the Landau-Lifshitz equation
\cite{usadel06prb,sukhovetal08j3m}. In the present work we use similar
methods (analytical and numerical) with the main objective here to
investigate the effects of surface anisotropy on the resonant absorption
by the spin-wave modes in box-shaped nanoclusters.

\subsubsection{\label{sub:LLE-SymplecticMethod}Solution of the Landau-Lifshitz
equation by symplectic methods}

In these numerical calculations we set $J=1$, $\hbar=1$. For simplicity,
we consider only cases in which the spins in the equilibrium state
are collinear and directed along the $z$ axis. This assumes that
the surface anisotropy does not exceed a certain critical value. Typically
we have $D=0.01$ and $D_{S}=0.1$. The ac field is applied along
the $x$ axis, if not stated otherwise. The results of this method
will be compared to those of the previous methods.

Among many existing solvers of systems of ordinary differential equations,
we employ a method making explicit rotations of spins around their
effective fields {[}see Refs. \cite{krechetal98cpc, stesch06cpc}
and many references therein{]}. This method conserves the spin length
and, in the absence of anisotropy, it also conserves the energy. Since
anisotropy is much weaker than the exchange interaction, the energy
non-conservation is weak. The evolution operator of the system corresponding
to the time interval $\Delta t$ can be written in the exponential
form 
\begin{equation}
\hat{U}=e^{\hat{L}\Delta t},\qquad\hat{L}=\sum_{i=1}^{\mathcal{N}}\hat{L}_{i}.
\end{equation}

There is no explicit formula for $e^{\hat{L}\Delta t}$ since the
precession of one spin changes the effective fields on the others.
However, the action of the operators $e^{\hat{L}_{i}\Delta t}$ describing
the rotation of an individual spin around its effective field with
all other spins frozen, can be worked out analytically. In the absence
of anisotropy this is simply the precession around a fixed field that
conserves both spin length and the energy. In the presence of anisotropy
the effective anisotropy field changes as the spin is precessing,
thus an analytical description of this precession is possible but
cumbersome. However, since the anisotropy field is much smaller than
the dominating exchange field, one can use the anisotropy field at
the beginning of the interval $\Delta t$, making only a small error.
Representing the precession of all spins in the system as a succession
of individual precessions induces errors growing with $\Delta t$.
This error can be reduced by using a generalization of the second-order
Suzuki-Trotter decomposition $e^{\left(\hat{A}+\hat{B}\right)h}=e^{\hat{A}h/2}e^{\hat{B}h}e^{\hat{A}h/2}+\mathcal{O}\left(h^{3}\right)$
that, in our case, has the form 
\begin{eqnarray}
\hat{U} & = & e^{\hat{L}_{1,}h}e^{\hat{L}_{2}h}\ldots e^{\hat{L}_{\mathcal{N}-1}h}e^{\hat{L}_{\mathcal{N}}h}e^{\hat{L}_{\mathcal{N}}h}e^{\hat{L}_{\mathcal{N}-1}h}\ldots e^{\hat{L}_{2}h}e^{\hat{L}_{1}h}
\end{eqnarray}
with $h\equiv\Delta t/2$. That is, all spins are rotated around their
respective effective fields in succession in some order. Then the
procedure is repeated in the reversed order. The effective field on
the next spin is updated because of rotation of the previous spin.
In the presence of a time-dependent field, the best choice is to take
the values of the latter in the middle of the two series of successive
rotations, that is, at $\Delta t/4$ and $3\Delta t/4$. Our implementation
of this method in Wolfram Mathematica (compiled) is rather efficient
and will be confirmed by agreement between the results obtained by
Eqs. (\ref{eq:Pabs_def}) and (\ref{eq:Pabs_via_ham}) for a not too
small time step, typically $\Delta t=0.1$.

We would like to emphasize that the approaches (analytical and numerical)
presented above are complementary and render the same results for
box-shaped clusters. However, the (numerical) method presented in
Section \ref{sub:LinLLE} is quite versatile as it allows us to compute
the excitation spectrum of a nanocluster of arbitrary shape and model
Hamiltonian.

\subsubsection{Definition and computing method of the absorbed power\label{sub:Absorbed_power_method}}

The power absorbed by a spin system in the presence of a uniform ac
magnetic field is defined as
\begin{equation}
P_{\mathrm{abs}}\left(t\right)=-\frac{1}{t_{f}}\intop_{0}^{t_{f}}dt\left(g\mu_{B}\right)\frac{1}{\mathcal{N}}\sum_{i}\left\langle \mathbf{s}_{i}\right\rangle \left(t\right)\cdot\frac{d\mathbf{H}_{\mathrm{ac}}(t)}{dt}\label{eq:Power3}
\end{equation}
where the integration is performed over time from the initial instant
$t=0$, at which all spins are in their (initial) equilibrium state,
to the final time $t_{f}$. Here, $\left\langle \mathbf{s}_{i}\right\rangle \left(t\right)\equiv\mathrm{Tr}\left[\rho(t)\mathbf{s}_{i}\right]$
where $\rho$ is the density matrix of the ferromagnet. Then, the
response of the spin system to a time-dependent field is defined by
the difference $\delta\left\langle s_{i}^{\alpha}\right\rangle (t)\equiv\left\langle s_{i}^{\alpha}\right\rangle \left(t\right)-\left\langle s_{i}^{\alpha}\right\rangle _{0}$,
with $\left\langle s_{i}^{\alpha}\right\rangle _{0}=\mathrm{Tr}\left(\rho_{0}\mathbf{s}_{i}\right)$,
$\rho_{0}$ being the density matrix of the unpurturbed ferromagnet.
However, in our calculations $t_{f}$ spans several periods, \emph{i.e.
$t_{f}=nT$} and as such, we can replace $\left\langle \mathbf{s}_{i}\right\rangle \left(t\right)$
by $\delta\left\langle \bm{s}_{i}\right\rangle (t)$ since the contribution
of the constant term vanishes. Therefore, the absorbed power becomes

\textcolor{black}{
\begin{equation}
P_{\mathrm{abs}}=-\frac{1}{t_{f}}\intop_{0}^{t_{f}}dt\left(g\mu_{B}\right)\frac{1}{\mathcal{N}}\sum_{i}\delta\left\langle \bm{s}_{i}\right\rangle (t)\cdot\dot{\mathbf{H}}_{\mathrm{ac}}\left(t\right).\label{eq:Pabs_def}
\end{equation}
}

On the other hand, since our model is conservative, the absorbed energy
should also be given by the change (per time) of the energy of the
system, leading to the equivalent definition 
\begin{equation}
P_{\mathrm{abs}}=\frac{1}{t_{f}\mathcal{N}}\left[\mathcal{H}(t_{f})-\mathcal{H}(0)\right].\label{eq:Pabs_via_ham}
\end{equation}
We use both formulae for the absorbed power that serve as a check
on the numerical calculations.

In order to clarify the expected form of the absorbed power that we
will compute numerically for magnetic nanoparticles, let us first
consider the simple case of a damped harmonic oscillator driven by
an oscillating force, \emph{i.e.} 
\begin{equation}
\ddot{x}+2\Gamma\dot{x}+\omega_{0}^{2}x=\xi h_{0}\sin\left(\omega t\right),\label{eq:Oscillator_Eq}
\end{equation}
where a coupling constant $\xi$ is introduced for generality. Solving
this equation with the initial conditions $x(0)=\dot{x}(0)=0$ and
calculating the absorbed power for times $t_{f}=N_{T}T$, $T=2\pi/\omega$,
$N_{T}$ being the number of cycles, with the help of Eq. (\ref{eq:Pabs_def})
{[}note that Eq. (\ref{eq:Pabs_via_ham}) cannot be used in the damped
case{]}, one obtains different results in different measurement time
ranges. At short times the result is that for the undamped harmonic
oscillator, 
\begin{equation}
\frac{P_{\mathrm{abs}}}{h_{0}^{2}}=\xi^{2}\frac{t_{f}}{2}\frac{1-\cos\left[\left(\omega-\omega_{0}\right)t_{f}\right]}{\left[\left(\omega-\omega_{0}\right)t_{f}\right]^{2}},\qquad\Gamma t_{f}\ll1.\label{eq:Pabs_oscill_undamped}
\end{equation}
The width of the corresponding peak decreases with the measurement
time as $\Delta\omega\sim1/t_{f}$, while its height grows linearly
with $t_{f}$, so that its integrated intensity is independent of
time. At long times a Lorentzian peak is formed around the (effective)
angular frequency \textcolor{black}{$\tilde{\omega}_{0}$} with 
\begin{equation}
\frac{P_{\mathrm{abs}}}{h_{0}^{2}}=\frac{\xi^{2}}{2}\frac{\Gamma}{\left(\omega-\tilde{\omega}_{0}\right)^{2}+\Gamma^{2}},\qquad\Gamma t_{f}\gg1.\label{eq:Pabs_oscill_damped}
\end{equation}

The latter formula is what is used in magnetic resonance experiments.
However, in numerical calculations on magnetic nanoparticles it is
inconvenient to perform a very long integration of the equations of
motion trying to measure damping that can be very small or zero. Eq.
(\ref{eq:Pabs_oscill_undamped}) that requires a relatively short
computation (we mainly use $N_{T}=10$) is fully sufficient in finding
the positions of resonance peaks and their intensities (parametrized
by the coupling constant $\xi$ in the oscillator model). In contrast
to the harmonic oscillator, SW modes in magnetic particles become
non-linear at high excitation thus leading to saturation and distortion
of the results. For this reason, in numerical calculations we have
to use the amplitude of the ac field $H_{0}$ as small as possible
without loss of precision in Eqs. (\ref{eq:Pabs_def}) and (\ref{eq:Pabs_via_ham}).

In the limit of a strong exchange coupling all spins are collinear
and can be considered as a single (macro-) spin with an effective
anisotropy stemming from the core and the surface. In this approximation,
the contribution of surface anisotropy is of first order in $D_{S}$
and depends on the particle's shape. For the case $D_{S}>0$ and oblate
particles in the $xy$ plane, the effective SA has an easy axis in
the $z$ direction. For prolate particles or for $D_{S}<0$ the $z$
direction becomes a hard axis of the effective SA. For particles of
cubic (or spherical) shape the first-order contribution of the SA
cancels out. However, there is a second-order contribution $\sim D_{S}^{2}/J$
that has a form of cubic anisotropy and which favours an orientation
of the particle's spin along the $(1,1,1)$ direction of the simple
cubic lattice. Indeed, this orientation leads to the largest deviations
from the collinear state that lower the total energy \cite{garkac03prl}.
Considering the precession of the macrospin (the particle's net magnetic
moment) in the effective field, to first order in $D_{S}$, one obtains
the resonance frequency 
\begin{eqnarray}
\hbar\omega_{0} & = & 2D\frac{\mathcal{N}_{\mathrm{core}}}{\mathcal{N}}\sqrt{\left[1+\frac{D_{S}}{D}\frac{N_{x}(N_{y}-N_{z})}{\mathcal{N}_{\mathrm{core}}}\right]\left[1+\frac{D_{S}}{D}\frac{N_{y}(N_{x}-N_{z})}{\mathcal{N}_{\mathrm{core}}}\right]}\label{eq:omega_0_first_order}
\end{eqnarray}
Indeed, for the cubic shape, $N_{x}=N_{y}=N_{z}$ and the effect of
$D_{S}$ vanishes. For $D_{S}>0$ and oblate particles ($N_{x},N_{y}>N_{z}$)
the resonance frequency increases, while for $D_{S}>0$ and prolate
particles the precession mode softens. We have not calculated the
second-order effect of SA on $\omega_{0}$ but the form of the effective
cubic anisotropy to second order in $D_{S}$ suggests that the precession
mode will soften for any sign of $D_{S}$, for the orientation of
spins along $z$ axis.

\section{Results and discussion}

\subsection{Surface and core contributions to the energy spectrum\label{sub:Surface-and-core-contribution}}

\begin{figure}[H]
\begin{centering}
\includegraphics[width=0.8\columnwidth]{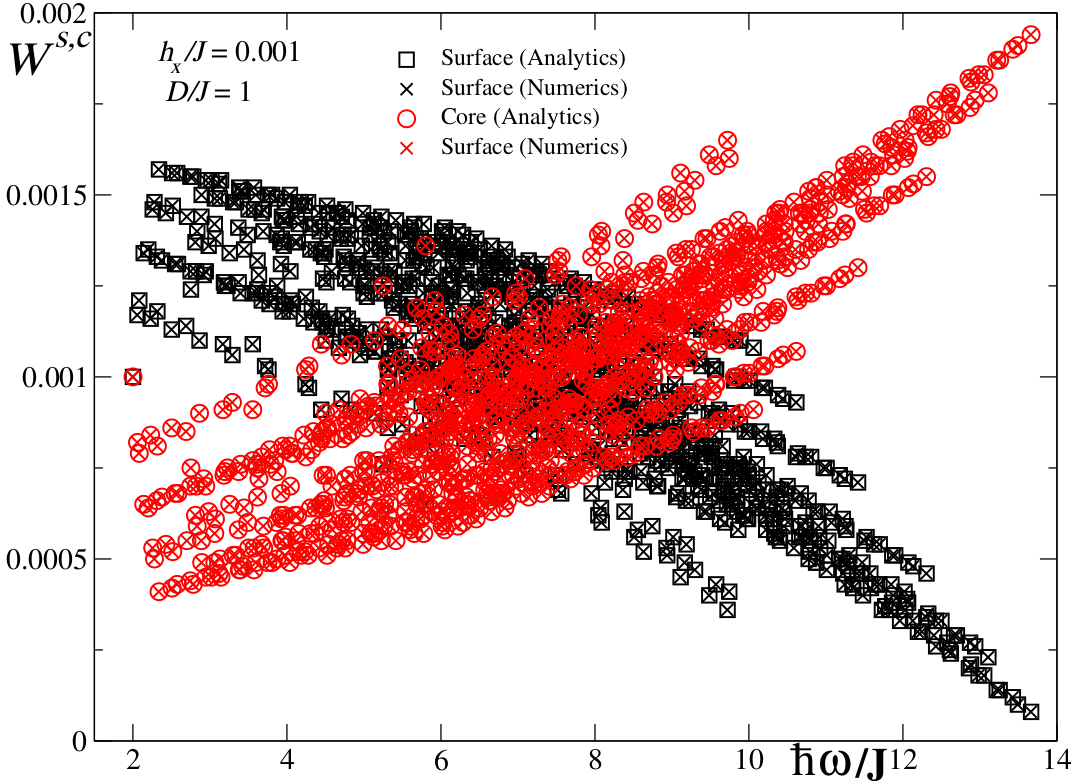} 
\par\end{centering}

\centering{}\protect\protect\caption{\label{fig:num_analy_cube}Spectral weight of spin-wave excitations
in a box-shaped particle of size $13\times11\times7$ and uniform
uniaxial anisotropy, in a magnetic field along the $x$ direction.}
\end{figure}

The procedure to determine the weight of surface and core spins in
the energy spectrum has been described in Subsection \ref{sub:LinLLE}.
In order to compare the spectral weights inferred from the analytical
expressions in Eq. (\ref{eq:Spinwave_function}) \emph{et seq} to
those obtained by the numerical method, we consider a box-shaped particle
with a simple cubic lattice. In order to avoid spurious effects that
could be due to highly symmetric systems we chose to investigate a
particle with sides of different lengths, \emph{e.g.} $N_{x}=13,\ N_{y}=11,\ N_{z}=7$.
In Fig. \ref{fig:num_analy_cube} we present a plot of the spectral
weight as a function of the energy $\hbar\omega$ (here $\hbar=1$)
in units of the nearest-neighbor exchange coupling $J$, with $J_{c}=J_{cs}=J_{s}=J$.
We have considered a static magnetic field along the $x$ axis and
a (uniform) uniaxial anisotropy for both the core and surface spins
with a common easy axis along the $z$ direction and anisotropy constant
$D/J=1$. The large core anisotropy $D=J$ is merely introduced in
order to shift the whole spectrum by $2J$ and thereby to highlight
the uniform mode. We can see that the numerical results fully agree
with the spectral weight inferred from the analytical eigenfunctions
in Eq. (\ref{eq:Spinwave_function}). The full spin-wave spectrum
of such many-spin systems is rather complex as it exhibits many branches,
and thence does not lend itself to a simple interpretation of the
various involved excitations. To that end, we have considered a representative,
though much simpler, system that consists of three coupled spin layers
for which the excitation spectrum can be computed, with the possibility
to disentangle the contributions of the surface and core layers. This
is done in \ref{sec:Toy-model}. The major difference is that the
three-layer toy model exhibits only three branches and we can see
that the surface spins dominate the low-frequency excitations. On
the other hand, the various branches of the many-spin system correspond
to different modes running in the $k-$space of a simple cubic lattice.
For instance, a quick inspection of Fig. \ref{fig:energy_spectrum_qxqy-1}
shows that the surface is dominant away from the Brillouin zone center.
In addition, the effect of the surface exchange coupling ($J_{s}$)
has been checked for the same particle without external magnetic field
or anisotropy. We have seen that at low excitation energies, the spectral
weights of the surface spins are always higher than those of the core
spins. However, as $J_{s}$ increases the branches of excitations
that preferentially involve surface spins merge with other branches
and thus decrease the surface contribution. This effect is more clearly
seen in the framework of the toy-model as shown in Fig. \ref{fig:energy_spectrum_qxqy-1}.

\subsection{Absorbed power }

\subsubsection{Box-shaped nanoparticles}

\begin{figure}
\begin{centering}
\includegraphics[width=0.48\textwidth]{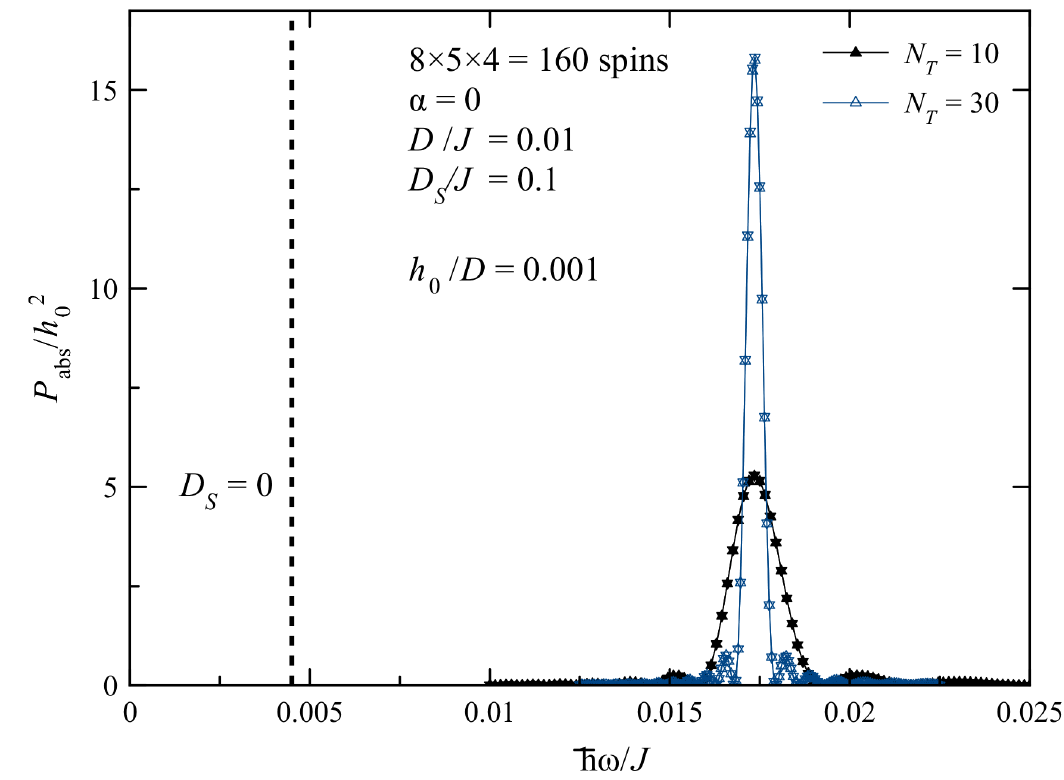}\includegraphics[width=0.48\textwidth]{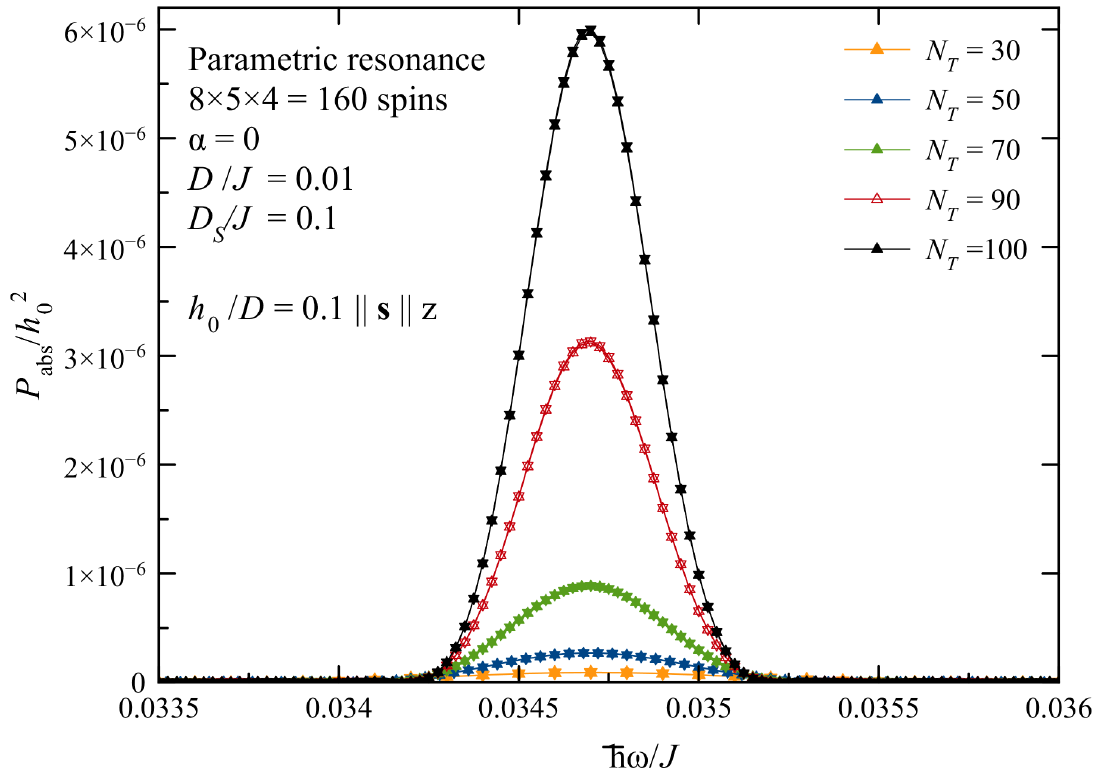}
\par\end{centering}

\centering{}\protect\caption{Absorbed power in a $8\times5\times4$ cubic particle. Left panel:
Magnetic resonance peak at $\hbar\omega/J=0.0174$ for two different
pumping times. The vertical dotted line shows the position of the
peak for $D_{S}=0$. Right panel: Parametric resonance peak at the
double frequency $\hbar\omega/J=0.0347$.\label{fig:8x5x4}}
\end{figure}

To see how our numerical method of Section \ref{sub:LLE-SymplecticMethod}
is implemented, we start with a small particle containing $160$ ($=8\times5\times4$)
spins that is flat in the $xy$ plane, with the anisotropy axis in
the $z$ direction and the ac field applied along the $x$ axis (if
not stated otherwise). The magnetic-resonance (MR) peak in Fig. \ref{fig:8x5x4}
(left panel) is seen at $\hbar\omega/J=0.0174$ that is far to the
right of the peak position $\hbar\omega/J=0.0045$ obtained for $D_{S}=0$.
This can be understood as the result of $xy$ planes having a larger
area, their stabilizing action for $D_{S}>0$ is stronger than the
destabilizing action of other surfaces, in a qualitative agreement
with Eq. (\ref{eq:omega_0_first_order}). One can see that increasing
the pumping time from $N_{T}=10$ to $N_{T}=30$ makes the resonance
peak narrower and higher, in accord with Eq. (\ref{eq:Pabs_oscill_undamped}).
Moreover, one can see the zeros of $P_{\mathrm{abs}}$ and small satellite
maxima between them. All the numerical work presented below uses $N_{T}=10$,
as this is sufficient to find the positions of the resonance maxima.
This is a shape effect indicating that the precession of spins is
elliptic rather than circular. In such cases parametric resonance
can be observed. Thus for the same particle, we also performed a parametric-resonance
calculation, directing the ac field in the spin direction $z$. The
results showing the initial stages of the exponential parametric instability
at the double frequency of the MR peak $\hbar\omega/J=0.0347$ are
shown in Fig. \ref{fig:8x5x4} (right panel). The parametric-resonance
peak has a different structure and its growth accelerates with the
pumping time. However, the parametric resonance requires a much stronger
amplitude of the ac field and longer pumping times, as compared with
MR peaks. In the sequel we will only concentrate on the latter.

\begin{figure}
\begin{centering}
\includegraphics[width=0.48\textwidth]{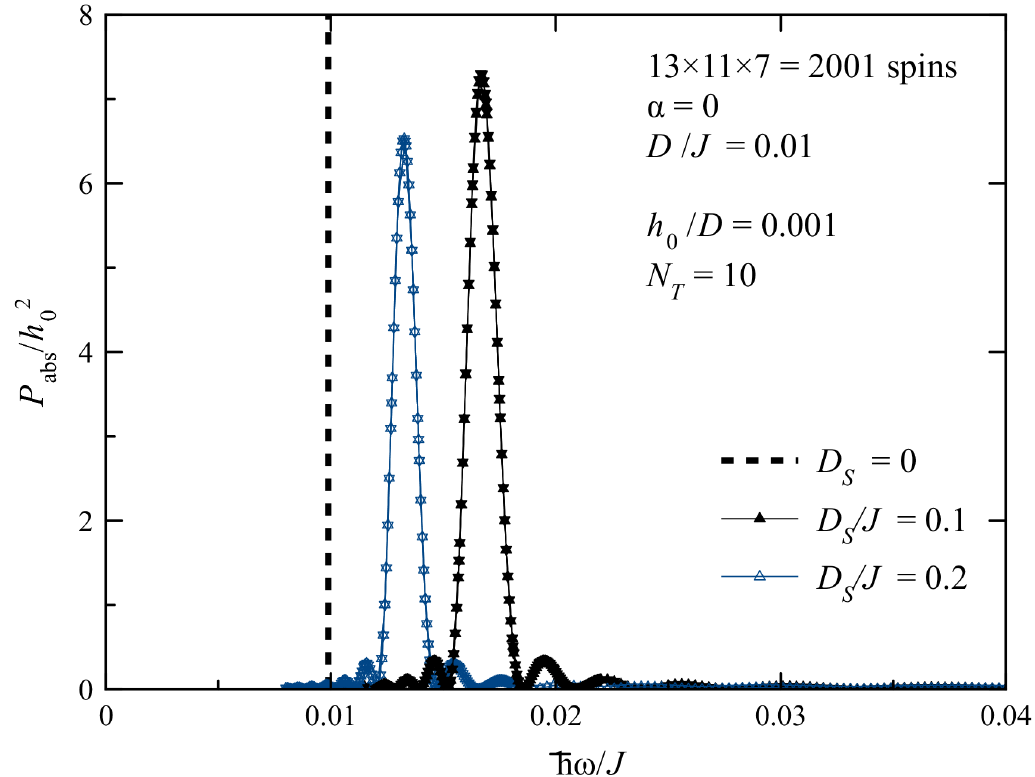} \includegraphics[width=0.48\textwidth]{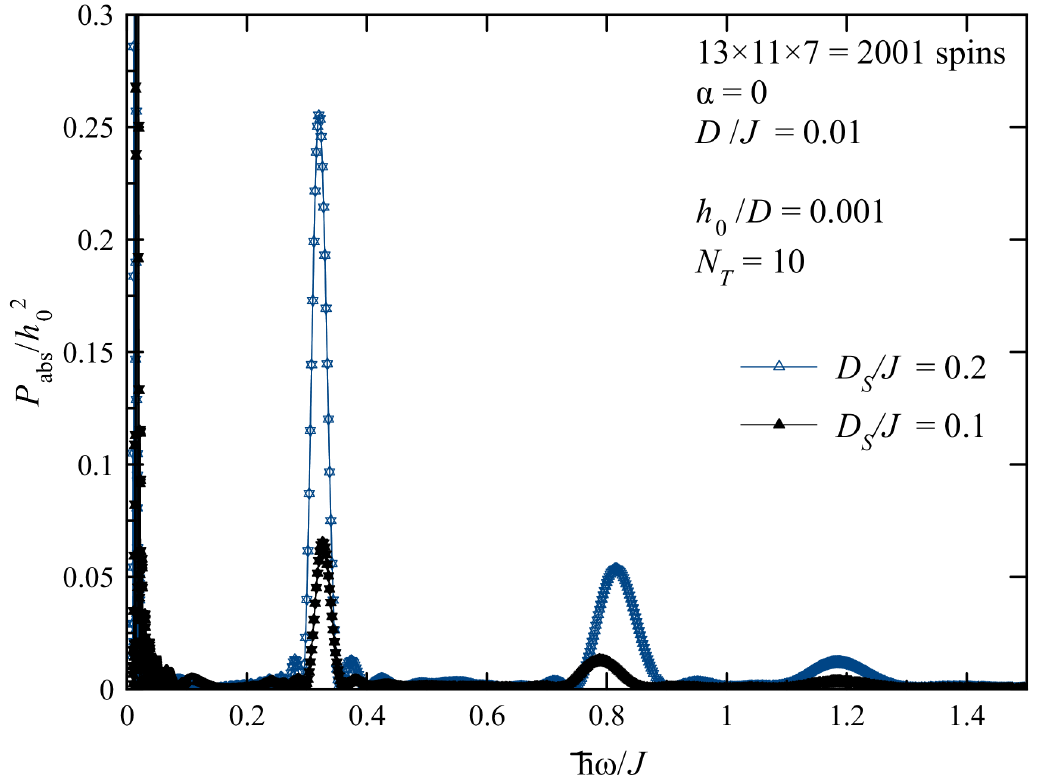} 
\par\end{centering}

\centering{}\protect\protect\caption{Absorbed power in a $13\times11\times7$ cubic particle. Left panel:
Low-frequency peak. The vertical dotted line shows the position of
the peak for $D_{S}=0$. Right panel: Both low-frequency peaks (far
left) and high-frequency peaks.\label{fig:13x11x7}}
\end{figure}

In order to identify the contributions from the core and surface spins
in the absorbed power we have investigated a cluster with a similar
aspect ratio as the cluster with $13\times11\times7=1001$ spins {[}see
Fig. \ref{fig:13x11x7}{]}, studied in Fig. \ref{fig:num_analy_cube}
and for which the diagonalization method presented in Section \ref{sub:LinLLE}
allows for a discrimination between the contributions from the core
and surface. 

\textcolor{black}{Taking the (space) Fourier transform of the spin
$s_{i}(t)$ in Eq. (\ref{eq:Pabs_def}) we obtain the power absorbed
by the $\bm{k}=\bm{0}$ mode}

\begin{eqnarray}
P_{\mathrm{abs}} & = & -\frac{1}{t_{f}}\intop_{0}^{t_{f}}dt\left(g\mu_{B}\right)\delta\bm{s}_{\bm{k=0}}(t)\cdot\dot{\mathbf{H}}_{\mathrm{ac}}\left(t\right)\nonumber \\
 & = & -\frac{1}{t_{f}}\intop_{0}^{t_{f}}dt\left(g\mu_{B}\right)\sum_{\alpha=x,y,z}\delta s_{\bm{k=0}}^{\alpha}(t){\bf e}_{\alpha}\cdot\dot{\mathbf{H}}_{\mathrm{ac}}\left(t\right).
\end{eqnarray}

\textcolor{black}{Then, setting $\bm{k}=\bm{0}$ in Eq. (\ref{eq:coeff_spins})}

\begin{eqnarray}
\delta\bm{s}_{\bm{k=0}}\left(t\right) & = & \delta s_{\bm{k=0}}^{\alpha}\left(0\right)e^{i\hbar\omega_{\bm{k}=\bm{0}}t}=\left(\sum_{j=1}^{\mathcal{N}}\sum_{\alpha=x,y,z}\mathcal{D}_{\bm{k}=\bm{0},j}^{\alpha}\bm{e}_{j}^{\alpha}\right)e^{i\hbar\omega_{\bm{k}=0}t}
\end{eqnarray}

\textcolor{black}{we obtain}

\begin{eqnarray}
P_{\mathrm{abs}} & = & -\frac{1}{t_{f}}\intop_{0}^{t_{f}}dte^{i\hbar\omega_{\bm{k}=\bm{0}}t}\left(\sum_{j=1}^{\mathcal{N}}\sum_{\alpha=x,y,z}\mathcal{D}_{\bm{k}=\bm{0},j}^{\alpha}\right)\left(\bm{e}_{j}^{\alpha}\cdot g\mu_{B}\dot{\mathbf{H}}_{\mathrm{ac}}\left(t\right)\right)
\end{eqnarray}

\textcolor{black}{Now, since the vectors ${\bf e}_{j}^{\alpha}$ are
all parallel to each other, i.e. ${\bf e}_{j}^{\alpha}={\bf e}^{\alpha}$,
the equation above simplifies into the following form}

\begin{eqnarray}
P_{\mathrm{abs}} & = & \left[-\frac{1}{t_{f}}\intop_{0}^{t_{f}}dte^{i\hbar\omega_{\bm{k}=\bm{0}}t}\left(g\mu_{B}\right)\left(\sum_{j=1}^{\mathcal{N}}\sum_{\alpha=x,y,z}\mathcal{D}_{\bm{k}=\bm{0},j}^{\alpha}\right)\right]\left(\bm{e}^{\alpha}\cdot\dot{\mathbf{H}}_{\mathrm{ac}}\left(t\right)\right)
\end{eqnarray}
\textcolor{black}{which suggests that we can introduce the power absorbed
by the degree of freedom (mode) corresponding to the component $\alpha=x,y,z$.
Indeed, we can write}

\begin{eqnarray}
P_{\mathrm{abs}}^{\alpha} & = & \left(\sum_{j=1}^{\mathcal{N}}\mathcal{D}_{\bm{k}=\bm{0},j}^{\alpha}\right)\times\left[-\frac{1}{t_{f}}\intop_{0}^{t_{f}}dte^{i\hbar\omega_{\bm{k}=\bm{0}}t}\bm{e}^{\alpha}\cdot g\mu_{B}\dot{\mathbf{H}}_{\mathrm{ac}}\left(t\right)\right].
\end{eqnarray}

This in turn can be rewritten as
\begin{equation}
P_{\mathrm{abs}}^{\alpha}\left(\bm{k}=0\right)=C_{\bm{k}=\bm{0}}^{\alpha}\times\widetilde{P}_{\bm{k}=\bm{0}}^{\alpha}\label{eq:Pabs_Coeff}
\end{equation}
\textcolor{black}{where 
\begin{equation}
C_{\bm{k}=\bm{0}}^{\alpha}\equiv\sum_{j=1}^{\mathcal{N}}\mathcal{D}_{\bm{k}=\bm{0},j}^{\alpha}\label{eq:coeff_Ck}
\end{equation}
is the statistical weight of the $\bm{k}=\bm{0}$ mode and 
\begin{equation}
\widetilde{P}_{\bm{k}=\bm{0}}^{\alpha}\equiv-\frac{1}{t_{f}}\intop_{0}^{t_{f}}dte^{i\hbar\omega_{\bm{k}=\bm{0}}t}\bm{e}^{\alpha}\cdot g\mu_{B}\dot{\mathbf{H}}_{\mathrm{ac}}\left(t\right).\label{eq:coeff_Pk}
\end{equation}
This means that the absorbed power (per mode) is proportional to the
sum of the coefficients of the wave-functions. As such, instead of
calculating the absorbed power as defined by Eq. (\ref{eq:Pabs_def})
we can calculate and plot the coefficients $C_{\bm{k}=\bm{0}}^{\alpha}$. }

\textcolor{black}{For a clearer analysis of the modes appearing in
the absorbed power spectrum, we first focus on the case of a box-shaped
sample with the same exchange constant everywhere, namely $J_{c}=J_{cs}=J_{s}=J$,
and without any anisotropy. All the spins are then identical and the
excitation spectrum is given by a single energy band in the $k$-space
as in Eq. (\ref{eq:Spinwave_function}). Hence, each mode can be unequivocally
labeled by its wave-vector $\bm{k}$ only.}\textcolor{blue}{{} }\textcolor{black}{According
to the definition of the coefficients $C_{k}^{\alpha}$, the power
can only be absorbed when the field couples to the uniform mode, i.e
for $C_{\bm{k}=\bm{0}}^{\alpha}=1$. On the other hand, for all other
values of the wave-vector $\mathbf{k}$ it can be easily shown that }

\begin{equation}
C_{\bm{k}\neq\bm{0}}^{\alpha}=\prod_{\alpha=x,y,z}\frac{\sin\left(N_{\alpha}k_{\alpha}\right)}{\sin\left(\frac{k_{\alpha}}{2}\right)}=0.
\end{equation}

\begin{figure}
\begin{centering}
\includegraphics[scale=0.33]{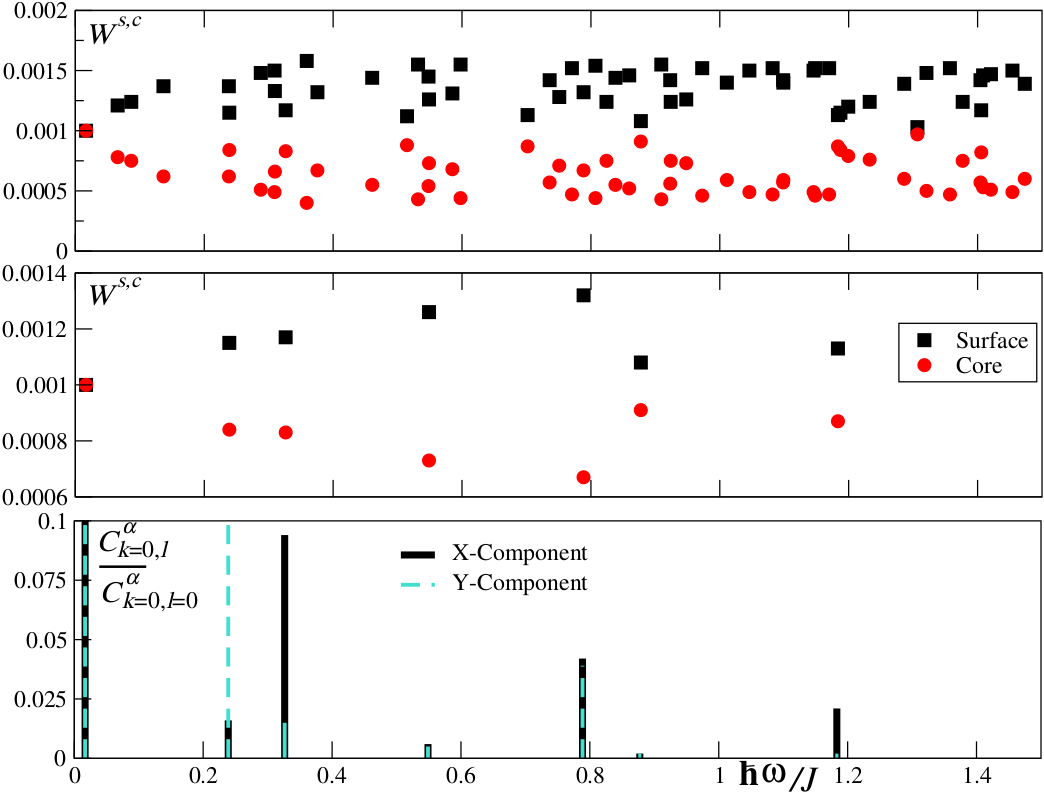}
\par\end{centering}

\centering{}\protect\caption{\label{fig:weight-power}Spectral weight of spin-wave excitations
in a box-shaped particle of size $13\times11\times7$. Upper panel
: spectral weight for the low frequency region. Middle panel: weights
for $C_{k=0,\ell}^{\alpha}\protect\neq0$. The lower panel correspond
to the coefficient $C_{k=0,\ell}^{\alpha}$ for the different components
of the spins defined in Eq. (\ref{eq:Pabs_Coeff}), normalized that
of the uniform mode $k=\ell=0.$}
\end{figure}

\textcolor{black}{In contrast to this simple case, for a system with
different types of local environments, as a consequence of an inhomogeneous
exchange coupling ($J_{c}\ne J_{cs}\ne J_{s}$), or of different types
of on-site anisotropies (surface and core), different energy bands
appear in the $\bm{k}$-space. This can be easily understood in the
framework of the toy model presented in \ref{sec:Toy-model} and shown
in Fig. \ref{fig:energy_spectrum_qxqy-1}). The analysis of such a
situation requires an additional band index ($\ell$) in order to
label each mode of energy $\hbar\omega_{k,\ell}$ and coefficients
$C_{k,\ell}^{\alpha}$. Consequently, the absorbed power can be attributed
to the non-uniform modes at $\bm{k}=\bm{0}$. This is shown in Fig.
\ref{fig:weight-power} which }presents the spectral weight and the
wave-function coefficients $C_{k,\ell}^{\alpha}$ in the low-frequency
regime, with a surface anisotropy $D_{S}/J=0.1$. The upper panel
shows the weights of the core and surface spins for all low frequency
modes.\textcolor{blue}{{} }\textcolor{black}{The middle and lower panels
respectively present the weights of the power-absorbing modes and
the coefficient $C_{k,\ell}^{\alpha}$, normalized by that of the
uniform mode ($k=\ell=0$).} We can see that the peaks in the absorbed
power in Fig. \ref{fig:13x11x7} (for $D_{S}/J=0.1$) coincide with
the peaks in black in Fig. \ref{fig:weight-power}, \emph{i.e} the
peaks obtained for an ac field applied along the $x$ axis. The peaks
in black obtained for $\hbar\omega/J=0.24,\ 0.54,\ 0.88$ in Fig.
\ref{fig:weight-power} are not seen in Fig. \ref{fig:13x11x7} because
the intensities of these peaks are too low compared to the satellites
obtained from the absorbed power, described in subsection \ref{sub:Absorbed_power_method}.
The first peak $\hbar\omega/J=0.017$ (in Fig. \ref{fig:13x11x7})
corresponds to the uniform mode. The latter corresponds to an equal
contribution (50\%) to the spectral weight from the core and surface
spins. Indeed, we have checked that this is in agreement with the
lowest energy mode shown in Fig. \ref{fig:weight-power} for which
the core and surface spectral weights coincide {[}see middle panel{]}.
Since the contribution of both core and surface spins is at its maximum
in this case, the low-energy peak in Fig. \ref{fig:13x11x7} and \ref{fig:weight-power}
exhibits the highest intensity\textcolor{black}{. The higher-frequency
peaks in black correspond to the non-uniform mode ($\bm{k}=0,\ \ell>0$)
due to the anisotropy and therefore they occur with a lower intensity.
These peaks have a dominant contribution from the surface spins (see
Tab. \ref{tab:table_weights}).}

\begin{table}
\begin{centering}
\begin{tabular}{|c|c|c|c|c|}
\hline 
$\hbar\omega/J$ & 0.017 & 0.33 & 0.79 & 1.18\tabularnewline
\hline 
\hline 
Surface (\%) & 50 & 60 & 70 & 60\tabularnewline
\hline 
Core (\%) & 50 & 40 & 30 & 40\tabularnewline
\hline 
\end{tabular}
\par\end{centering}

\centering{}\protect\caption{\label{tab:table_weights}Contributions to the spectral weight from
the surface and core spins for the cluster $13\times11\times7$ with
$D_{S}/J=0.1$ and a time-dependent field applied along the $x$ axis
(i. e. Black peaks of Fig.\ref{fig:weight-power}).}
\end{table}

The peaks in cyan in Fig.\ref{fig:weight-power} are obtained for
a time-dependent field along the $y$ axis. These peaks appear with
the same frequencies as the peaks in black but with different intensities.
In addition, the contributions from the surface and core spins may
vary from one type of peaks to the other.

For the same nanocluster and in accordance with Eq. (\ref{eq:omega_0_first_order}),
in Fig. \ref{fig:13x11x7} the position of the low-frequency peak
shifts to the right as $D_{S}$ increases from zero (compare with
the vertical line at $D_{S}=0$). However, a further increase of SA
reverses this tendency, as can be seen from the curve $D_{S}/J=0.2$.
This mode softening can be attributed to the second-order effect of
surface anisotropy.\textcolor{black}{{} On the other hand, in the high-frequency
part of the spectrum one can observe three peaks that could be attributed
to three different types of the nanocluster facets with different
local environment (or effective fields). Note that the positions of
the peaks are nearly the same for $D_{S}/J=0.1$ and $D_{S}/J=0.2$,
which hints at the predominant exchange origin of these modes.}

\subsubsection{Size effect and application to nanocubes}

\begin{figure*}
\begin{centering}
\includegraphics[height=5cm]{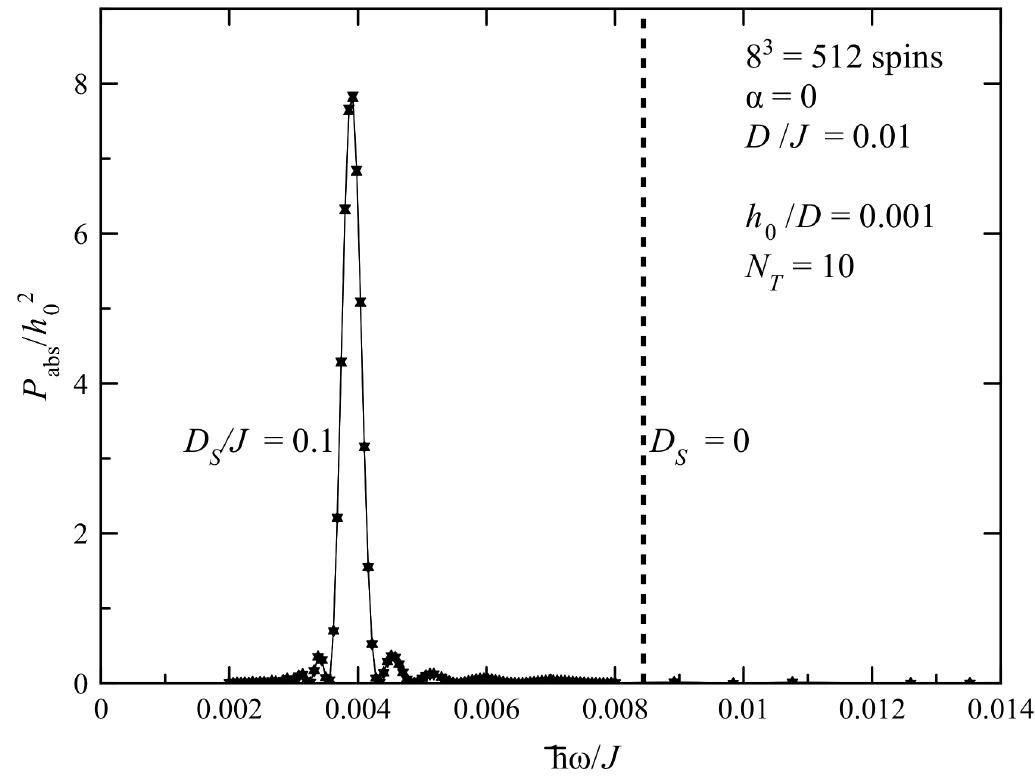}\includegraphics[height=5cm]{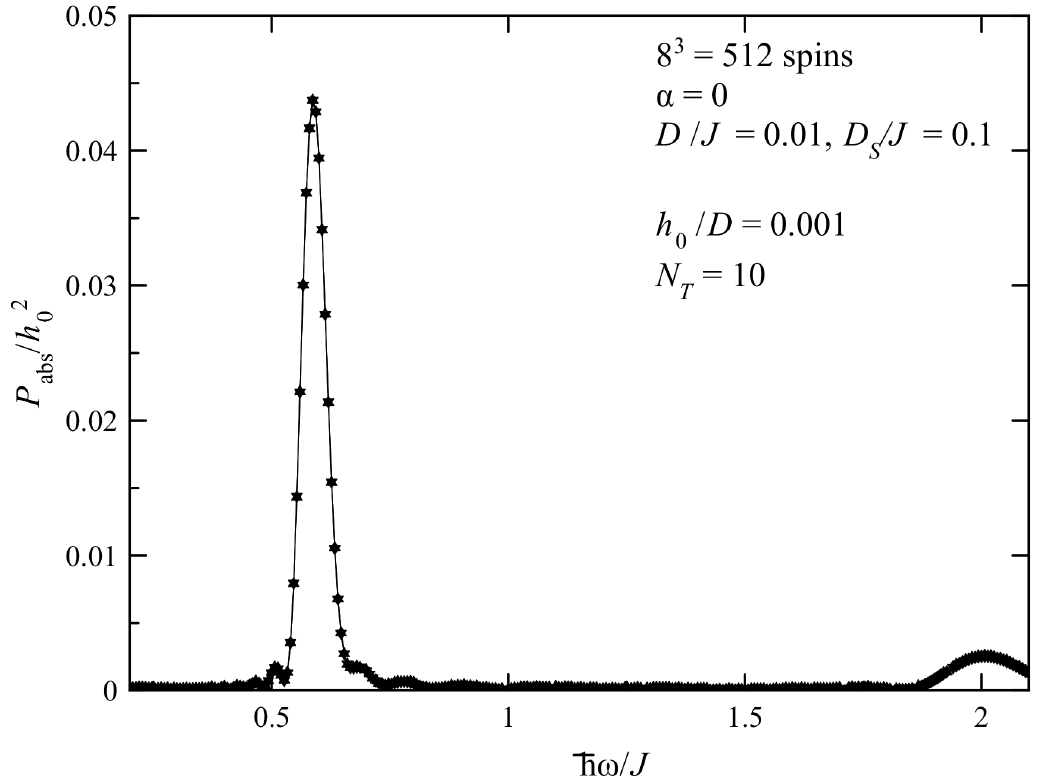} 
\par\end{centering}

\begin{centering}
\includegraphics[height=5cm]{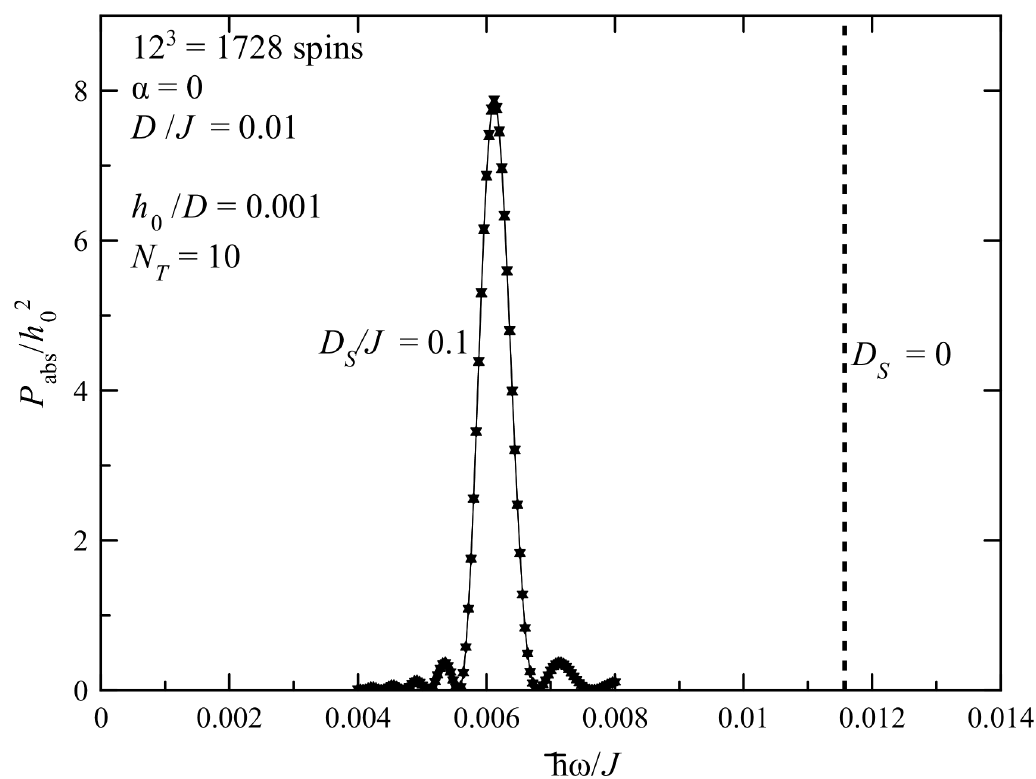}\includegraphics[height=5cm]{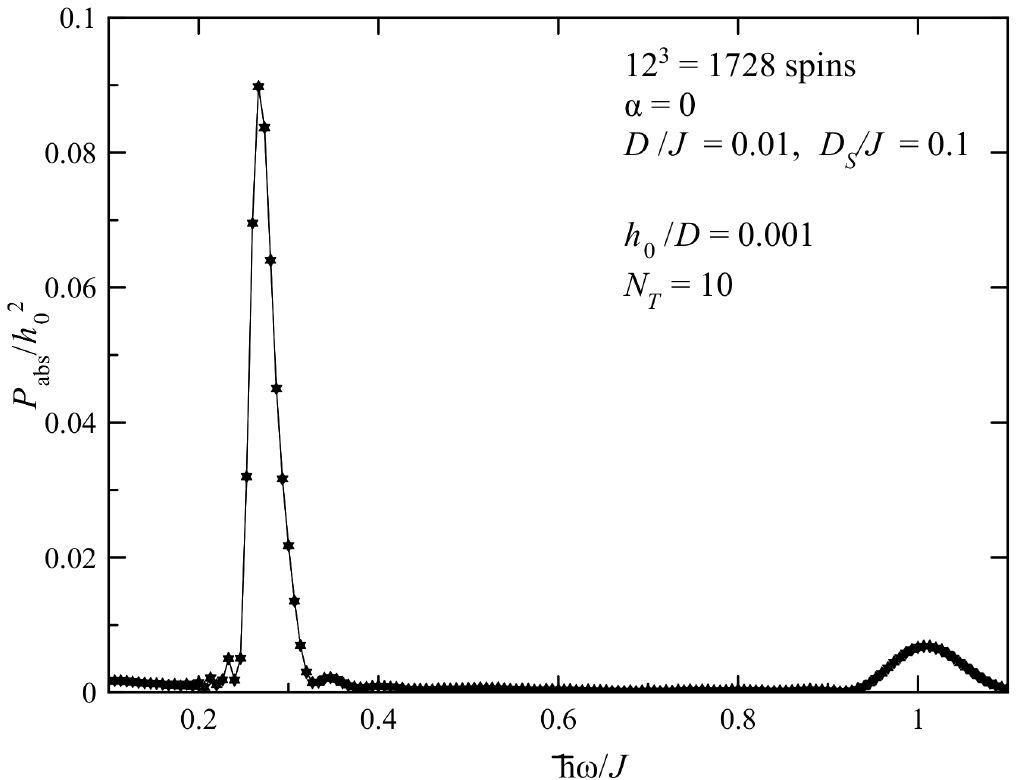} 
\par\end{centering}

\centering{}\protect\protect\caption{Absorbed power for $8\times8\times8$ and $12\times12\times12$ particles
with a focus on the low-frequency peaks in the left column and the
high-frequency peaks in the right column. The vertical dotted line
shows the position of the peak for $D_{S}=0$.\label{fig:cubes}}
\end{figure*}

The investigation of size effects in general (\emph{i.e.} without
any rotational symmetry) is a rather involved task since upon increasing
the size the number of modes increases and their degeneracy makes
it difficult to disentangle their contributions to the spectral weight.
This is one of the reasons for which we have decided to focus on cubic
samples. In fact, today samples of (iron) nanocubes are routinely
investigated in experiments since their synthesis has become fairly
well controlled. 

Accordingly, the results for the absorbed power for the $8\times8\times8$
particle (512 spins) are shown in Fig. \ref{fig:cubes}. One can see
a strong peak at $\hbar\omega=0.0039J$ that corresponds to nearly
coherent precession of all spins in the particle. Because of the second-order
effect of SA \cite{garkac03prl} this peak is shifted to the left
from its position for $D_{S}=0$, shown by the vertical dotted line
at $\hbar\omega_{0}=2D\mathcal{N}_{\mathrm{core}}/\mathcal{N}=0.0084J$.
Note that the first-order formula, Eq. (\ref{eq:omega_0_first_order}),
does not capture this effect. Here one cannot use $D_{S}/J=0.2$ because
further shift of the peak to the left renders the collinear spin configuration
along the $z$ axis unstable.

\begin{figure}
\begin{centering}
\includegraphics[width=0.7\columnwidth]{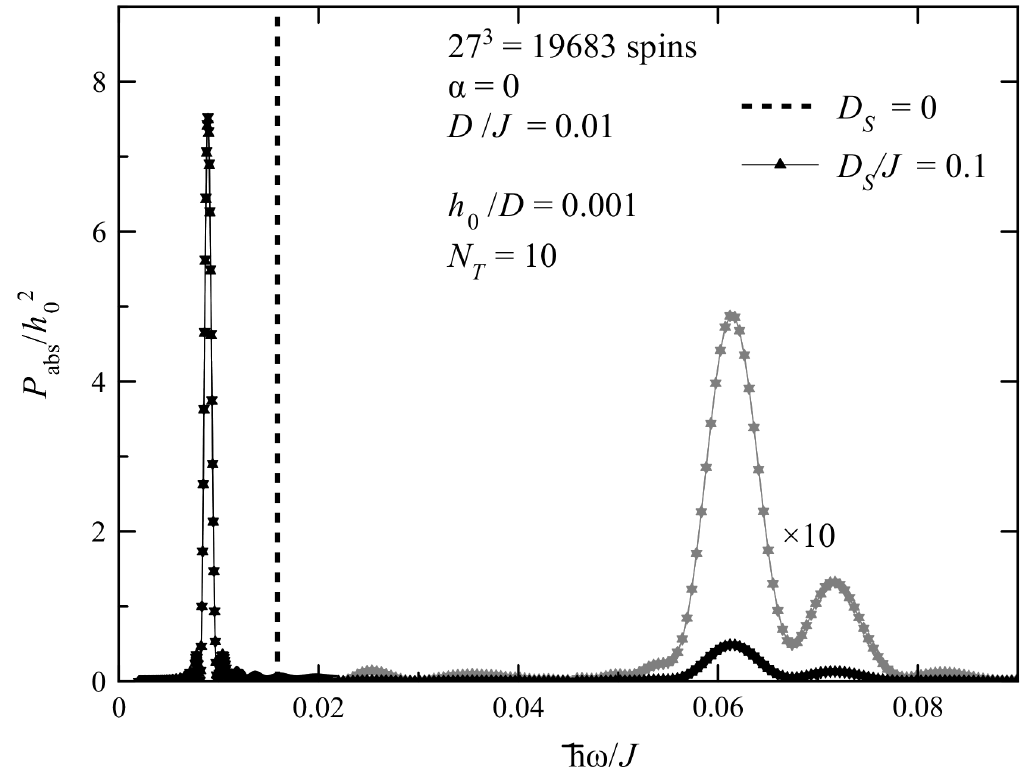} 
\par\end{centering}

\protect\caption{Same as in Fig. \ref{fig:cubes} but for the cluster $27\times27\times27$.\label{fig:FeNanocubes}}
\end{figure}

The lower panel of Fig. \ref{fig:cubes} shows similar results for
a larger particle of $12\times12\times12=1728$ spins. Here the low-frequency
peak is shifted to the right in comparison with the $8\times8\times8$
particle, and which can be explained by the smaller fraction of surface
spins. The leftmost and strongest of high-frequency peaks here is
larger and shifted to the left. Note that for both of these sizes
high-frequency peaks are much smaller than the main low-frequency
peak (notice the difference in scale between the left and right panels).
By way of illustration, we consider an Fe nanocube of side $a=8\,\mathrm{nm}$
\cite{trunovaetal08jap,Respaud_nanocube2008,Respaud_nanocube2_2010JMMM,jiangetal10jac,kronastetal11nl,okellyetal12nanotech}.
This corresponds to a nanocluster of size $27\times27\times27$ particle
whose absorption spectrum is shown in Fig. \ref{fig:FeNanocubes}.

Although the present paper is focused on theoretical aspects, a few
predictions can be made for realistic iron nanocubes studied today
in many experiments. Both synthesis and recent experimental developments
have provided systems with optimized structures that could be mimicked
by the simplified model studied here. In particular, using some oxygen
and plasma treatment it seems that the ligands and oxide shell could
be effectively removed, leaving us with ferromagnetic nanocubes, see
\emph{e.g.} Ref. \cite{trunovaetal08jap}. Would FMR measurements
on such nanocubes become possible, the observed spectrum should exhibit
the features described in the present work, \emph{e.g.} a low-energy
peak at around $10{\rm \ GHz}$ for the uniform mode, followed by
higher-energy excitations that couple to the latter. In addition,
the aspect-ratio of box-shaped (non-cubic) samples can be figured
out by this technique upon checking whether a parametric resonance
feature appears in the spectrum. In regards with the values of the
physical parameters taken in our calculations, we note that the ratio
of the magneto-crystalline anisotropy to the exchange coupling ($D/J$)
is here taken at least an order of magnitude larger than in typical
iron systems. The reason is that lower values of this ratio require
much more time-consuming calculations while the physical picture remains
the same. More precisely, the calculation for typical iron materials
with $D/J\sim10^{-3}$ would lead to a down-shift of the low-frequency
peak roughly by a factor of 10 while the high-frequency peaks should
practically remain the same.

As compared with the sizes dealt with above, here the high-frequency
peak is even larger and even more shifted to the left, so that the
low- and high-frequency spectra can be plotted on the same graph.
In addition, the high-frequency peak resolves into two peaks. The
main high-frequency peak, shown in the right column of Fig. \ref{fig:cubes},
can be interpreted as being due to the precession of the spins located
near the facets of the cube. Since this precession mode is non-uniform
(it has a non-zero $k=0$ component) there is exchange energy involved
and this is why the precession frequency is high. With an increasing
size, the exchange energy per spin in this mode decreases, and so
does its frequency. The splitting of the main high-frequency peak
seen for the $27\times27\times27$ particle can be explained by the
fact that SA induces an increase of the mode stiffness at the two
$xy$ planes (the small peak on the right) and to a decrease of the
mode stiffness at the four other surfaces (the big peak on the left).

\section{Conclusion}

Through a systematic numerical investigation, backed by analytical
calculations for special cases, we have studied and distinguished
the role of surface and core spins in box-shaped magnetic nanoparticles.
We have focused this work on this specific shape inspired by numerous
experimental studies of iron nanocubes which are now available in
well controlled cubic shapes and sizes. On the other hand, ferromagnetic
resonance measurements on ``isolated'' nanoelements has now become
possible with the necessary sensitivity for measuring the absorbed
power. 

Accordingly, we have computed the absorbed power as a function of
the excitation frequency and have shown that it is possible to attribute
the different contributions of the surface and those of the core spins
to the various peaks obtained in our calculations. In particular,
the low-energy peak, corresponding to the $\bm{k=0}$ mode, consists
of equal contributions from the surface and core spins. Furthermore,
in the case of less symmetric box-shaped samples with Néel surface
anisotropy, we observe an elliptic precession of the spins whose signature
can be seen in a parametric resonance experiment, where a small signal
should be detected at twice the frequency of the standard magnetic
resonance response.

\ack{}{This work was partly supported by the French ANR/JC MARVEL. The work
of DG was also supported by the U.S. National Science Foundation through
Grant No. DMR-1161571. }

\appendix

\section{\label{sec:Spherical-coordinates}Energy Hessian in spherical coordinates}

At each site $i$ of the cluster's lattice we may define the reference
system with the spherical coordinate $(\theta_{i},\varphi_{i})$ and
basis basis $(\mathbf{s}_{i},\mathbf{e}_{\theta_{i}},\mathbf{e}_{\varphi_{i}})$
related to the Cartesian coordinates by

\begin{eqnarray}
\mathbf{s}_{i} & = & \mathbf{e}_{\varphi_{i}}\left(\begin{array}{c}
\sin\theta_{i}\cos\varphi_{i}\\
\sin\theta_{i}\sin\varphi_{i}\\
\cos\theta_{i}
\end{array}\right),\ \mathbf{e}_{\theta_{i}}=\left(\begin{array}{c}
\cos\theta_{i}\cos\varphi_{i}\\
\cos\theta_{i}\sin\varphi_{i}\\
-\sin\theta_{i}
\end{array}\right),\ \mathbf{e}_{\varphi_{i}}=\left(\begin{array}{c}
-\sin\varphi_{i}\\
\cos\varphi_{i}\\
0
\end{array}\right).\label{eq:SCUnitVectors}
\end{eqnarray}

From this we derive 
\begin{eqnarray*}
\partial_{\theta_{i}}\mathbf{s}_{i} & = & \mathbf{e}_{\theta_{i}},\quad\partial_{\varphi_{i}}\mathbf{s}_{i}\mathbf{=}\sin\theta_{i}\,\mathbf{e}_{\varphi_{i}},\\
\partial_{\theta_{i}}\mathbf{e}_{\theta_{i}} & = & -\mathbf{s}_{i}\mathbf{,\quad}\partial_{\varphi_{i}}\mathbf{e}_{\theta_{i}}=\cos\theta_{i}\,\mathbf{e}_{\varphi_{i}},\\
\partial_{\theta_{i}}\mathbf{e}_{\varphi_{i}} & = & 0,\;\partial_{\varphi_{i}}\mathbf{e}_{\varphi_{i}}=-\left(\sin\theta_{i}\,\mathbf{s}_{i}+\cos\theta_{i}\,\mathbf{e}_{\theta_{i}}\right).
\end{eqnarray*}
leading to $\delta\mathbf{s}_{i}=\delta\theta_{i}\partial_{\theta_{i}}\mathbf{s}_{i}\mathbf{+}\delta\varphi_{i}\partial_{\varphi_{i}}\mathbf{s}_{i}\mathbf{=}\delta\theta_{i}\mathbf{e}_{\theta_{i}}\mathbf{+}\delta\varphi_{i}\sin\theta_{i}\,\mathbf{e}_{\varphi_{i}}$.
Then using the gradient

\begin{equation}
\partial_{\mathbf{s}_{i}}\equiv\mathbf{\bm{\nabla}}_{i}=\mathbf{e}_{\theta_{i}}\partial_{\theta_{i}}+\mathbf{e}_{\varphi_{i}}\frac{1}{\sin\theta_{i}}\,\partial_{\varphi_{i}}\mathbf{,}
\end{equation}
we get $\delta\mathbf{s}_{i}\mathbf{\cdot\bm{\nabla}}_{i}=\delta\theta_{i}\,\partial_{\theta_{i}}+\delta\varphi_{i}\,\partial_{\varphi_{i}}$.
This implies for an arbitrary function $f(\theta_{i}\varphi_{i})$
\begin{equation}
\partial_{\theta_{i}}f=\mathbf{e}_{\theta_{i}}\cdot\mathbf{\bm{\nabla}}_{i}f,\quad\partial_{\varphi_{i}}f=\sin\theta_{i}\,\mathbf{e}_{\varphi_{i}}\cdot\mathbf{\bm{\nabla}}_{i}f.
\end{equation}

Since the spin deviation $\delta\bm{s}_{k}$ can be written in terms
of $\delta\theta_{k}$ and $\delta\varphi_{k}$ Eq. (\ref{eq:eigenvalue_pb})
can be written in the basis $\left\{ \left({\bf e}_{\theta_{i}},{\bf e}_{\varphi_{i}}\right)\right\} _{i=1,\cdots,\mathcal{N}}=\left\{ \bm{\xi}_{\mu}\right\} _{\mu=1,\cdots,2\mathcal{N}}$.
Note, however, that in the general case these unit vectors are not
orthogonal to each other \emph{i.e.} $\bm{\xi}_{\mu}\cdot\bm{\xi}_{\nu}\ne\delta_{\mu,\nu}$.
In fact, $\delta\bm{s}_{k}$ represents the usual spin-wave deviations
from the local equilibrium state of spin ${\bf s}_{k}$, which is
denoted by ${\bf s}_{k}^{\left(0\right)}$. The latter represents
the quantization direction for the local algebra. It's well known
that $\delta{\bf s}_{k}$ can be written in terms of the spin operators
$S_{k}^{\pm}$ which form a local SU(2) algebra with the usual commutation
rules, \emph{i.~e.} $\left[S_{i}^{\alpha},S_{j}^{\beta}\right]=i\varepsilon^{\alpha\beta\gamma}\delta_{ij}S_{i}^{\gamma}$,
with $\varepsilon^{\alpha\beta\gamma}$ being the Levi-Civita tensor.
In particular, spins operating on different sites commute with each
other. This implies that the vectors $\delta{\bf s}_{k}$, or more
precisely, the transverse vectors $\left\{ \left({\bf e}_{\theta_{i}},{\bf e}_{\varphi_{i}}\right)\right\} _{i=1,\cdots,\mathcal{N}}=\left\{ \bm{\xi}_{\mu}\right\} _{\mu=1,\cdots,2\mathcal{N}}$
can be represented by the vectors \textcolor{black}{of the orthonormal
canonical basis $\left\{ \left({\bf e}_{i}\right)\right\} _{i=1,\cdots,\mathcal{N}}$
with $e_{i}^{\alpha}=\delta_{i,\alpha}$.}

Assuming that the energy $\mathcal{E}=\sum_{i=1}^{\mathcal{N}}\mathcal{E}_{i},$
is given by a general Hamiltonian we obtain the second derivatives
of $\mathcal{E}_{i}$ in terms of its derivative with respect to $\mathbf{s}_{i}$.
\begin{equation}
\bm{\widetilde{\mathcal{H}}}_{ik}(\mathcal{E}_{i})\equiv\left(\begin{array}{ll}
\partial_{\theta_{i}\theta_{k}}^{2}\mathcal{E} & \frac{1}{\sin\theta_{i}}\partial_{\theta_{k}\varphi_{i}}^{2}\mathcal{E}\\
\\
\frac{1}{\sin\theta_{k}}\partial_{\varphi_{k}\theta_{i}}^{2}\mathcal{E} & \frac{1}{\sin\theta_{i}\sin\theta_{k}}\partial_{\varphi_{k}\varphi_{i}}^{2}\mathcal{E}
\end{array}\right).\label{eq:SC-pseudoHessian}
\end{equation}

This is the (pseudo-) Hessian of $\mathcal{E}$ resulting from the
action of the (pseudo-) Hessian operator 
\begin{equation}
\bm{\widetilde{\mathcal{H}}}_{ik}=\mathbf{\bm{\nabla}}_{k}^{T}\mathbf{\bm{\nabla}}_{i}=\left(\begin{array}{cc}
\partial_{\theta_{k}} & \frac{1}{\sin\theta_{k}}\partial_{\varphi_{k}}\end{array}\right)\left(\begin{array}{c}
\partial_{\theta_{i}}\\
\\
\frac{1}{\sin\theta_{i}}\partial_{\varphi_{i}}
\end{array}\right).
\end{equation}

For a given nanocluster of given size, shape, anisotropy model and
the applied field, one first determines the equilibrium state, denoted
by $\left\{ {\bf s}_{i}^{(0)}=(\theta_{i}^{(0)},\varphi_{i}^{(0)})\right\} _{i=1,\cdots,\mathcal{N}}$,
where $\theta_{i}$ and $\varphi_{i}$ are the standard spherical
angles defined with respect to the local basis $\left({\bf s}_{i},\mathbf{e}_{\theta_{i}},\mathbf{e}_{\varphi_{i}}\right)$
at site $i$.

The effective field is defined by $\mathbf{H}_{\mathrm{eff},i}=-\delta_{\mathbf{s}_{i}}\mathcal{E}=-\mathbf{\bm{\nabla}}_{i}\mathcal{E}$,
such that the four second derivatives read 
\begin{eqnarray}
\partial_{\theta_{i}\theta_{k}}^{2}\mathcal{E} & = & \delta_{ik}\left[\mathbf{s}_{i}\cdot-\mathbf{e}_{\theta_{i}}\cdot\left(\mathbf{e}_{\theta_{i}}\cdot\mathbf{\bm{\nabla}}_{i}\right)\right]\mathbf{H}_{\mathrm{eff},i}-\left(1-\delta_{ik}\right)\mathbf{e}_{\theta_{i}}\cdot\left[\mathbf{e}_{\theta_{k}}\cdot\mathbf{\bm{\nabla}}_{k}\right]{\bf H}_{\mathrm{eff},i},\nonumber \\
\nonumber \\
\partial_{\varphi_{k}\varphi_{i}}^{2}\mathcal{E} & = & \delta_{ik}\sin\theta_{i}\left[\left(\sin\theta_{i}\,\mathbf{s}_{i}+\cos\theta_{i}\,\mathbf{e}_{\theta_{i}}\right)-\sin\theta_{i}\,\mathbf{e}_{\varphi_{i}}\cdot\left(\mathbf{e}_{\varphi_{i}}\cdot\mathbf{\bm{\nabla}}_{i}\right)\right]\mathbf{H}_{\mathrm{eff},i}\nonumber \\
 &  & -\left(1-\delta_{ik}\right)\sin\theta_{i}\sin\theta_{k}\,\mathbf{e}_{\varphi_{i}}\cdot\left[\mathbf{e}_{\varphi_{k}}\cdot\mathbf{\bm{\nabla}}_{k}\right]\mathbf{H}_{\mathrm{eff},i},\nonumber \\
\nonumber \\
\partial_{\theta_{k}\varphi_{i}}^{2}\mathcal{E} & = & -\delta_{ik}\left[\cos\theta_{i}\,\mathbf{e}_{\varphi_{i}}\cdot+\sin\theta_{i}\,\mathbf{e}_{\varphi_{i}}\cdot\left(\mathbf{e}_{\theta_{i}}\cdot\mathbf{\bm{\nabla}}_{i}\right)\right]\mathbf{H}_{\mathrm{eff},i}\nonumber \\
 &  & -\left(1-\delta_{ik}\right)\sin\theta_{i}\,\mathbf{e}_{\varphi_{i}}\cdot\left[\mathbf{e}_{\theta_{k}}\cdot\mathbf{\bm{\nabla}}_{k}\right]\mathbf{H}_{\mathrm{eff},i},\nonumber \\
\nonumber \\
\partial_{\varphi_{k}\theta_{i}}^{2}\mathcal{E} & = & -\delta_{ik}\left[\cos\theta_{i}\,\mathbf{e}_{\varphi_{i}}\cdot+\sin\theta_{i}\,\mathbf{e}_{\theta_{i}}\cdot\left(\mathbf{e}_{\varphi_{i}}\cdot\mathbf{\bm{\nabla}}_{i}\right)\right]\mathbf{H}_{\mathrm{eff},i}\nonumber \\
 &  & -\left(1-\delta_{ik}\right)\sin\theta_{k}\,\mathbf{e}_{\theta_{i}}\cdot\left(\mathbf{e}_{\varphi_{k}}\cdot\mathbf{\bm{\nabla}}_{k}\right)\mathbf{H}_{\mathrm{eff},i}.\label{eq:Energy2Derivatives}
\end{eqnarray}

It is understood that all these derivatives and the pseudo-Hessian
have to be evaluated at the equilibrium state $\left\{ {\bf s}_{i}^{(0)}=(\theta_{i}^{(0)},\varphi_{i}^{(0)})\right\} _{i=1,\cdots,\mathcal{N}}$.

\section{Toy model\label{sec:Toy-model}}

\begin{figure}[H]
\begin{centering}
\includegraphics[width=0.5\columnwidth]{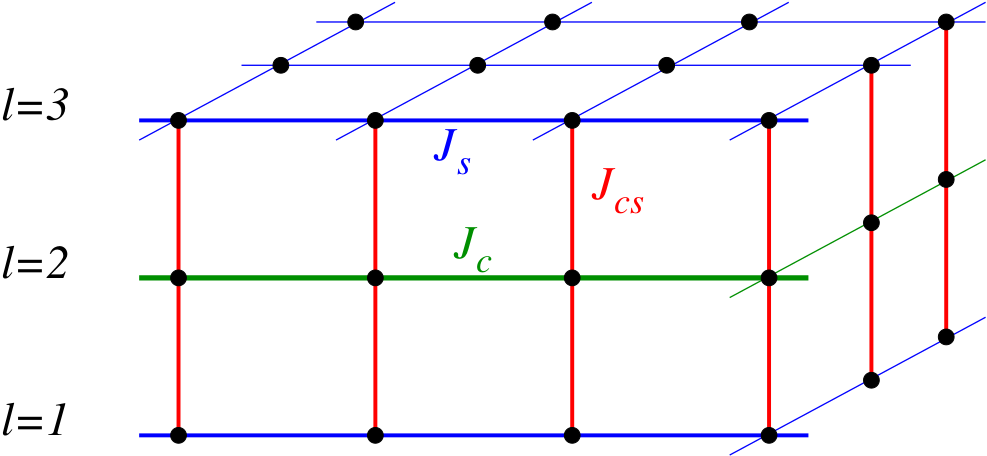} 
\par\end{centering}

\centering{}\protect\caption{\label{fig:sample} $2D$ Slab of $3$ atomic layers with exchange
couplings $J_{s}$ (surface), $J_{cs}$ (core-surface) and $J_{c}$
(core). }
\end{figure}

In order to achieve a simple physical picture of the contributions
of core and surface spins to the spectral weight, together with a
possible comparison with the numerical method developed in Subsection
\ref{sub:LinLLE}, we have built a toy model that captures the main
feature we want to illustrate but which is analytically tractable.
Accordingly, we consider a ferromagnet composed of 3 coupled layers
as sketched in Fig. \ref{fig:sample}. Each layer is assumed to be
infinite in $x$ and $y$ directions.

The spin Hamiltonian of such a system is the Heisenberg model 
\begin{equation}
\begin{array}{lll}
\mathcal{H} & = & -{\displaystyle \sum_{l=1,3}}J_{l}{\displaystyle \sum_{i}}\bm{S}_{i,l}\cdot\left(\bm{S}_{i+x,l}+\bm{S}_{i+y,l}\right)-J_{cs}{\displaystyle \sum_{i}}\bm{S}_{i,2}\cdot\left(\bm{S}_{i,1}+\bm{S}_{i,3}\right)\end{array},\label{eq:hamiltonian-heisenberg}
\end{equation}
where $\bm{S}_{i,l}$ is the spin at site $i$ within the layer $l$,
and $J_{l=1,3}\equiv J_{s}$ and $J_{2}\equiv J_{c}$. We restrict
ourselves to the case of a ferromagnet with $J_{l}>0,J_{cs}>0$. In
the spin-wave approach we choose $z$ as the quantization axis and
perform a Holstein-Primakoff transformation

\begin{eqnarray}
S_{i,l}^{z} & = & S_{i,l}^{-}S-a_{i,l}^{\dagger}a_{i,l},\qquad S_{i,l}^{+}\simeq\sqrt{2S}a_{i,l},\qquad S_{i,l}^{-}\simeq\sqrt{2S}a_{i,l}^{\dagger}.
\end{eqnarray}

\begin{figure*}
\noindent \begin{centering}
\includegraphics[height=4.6cm]{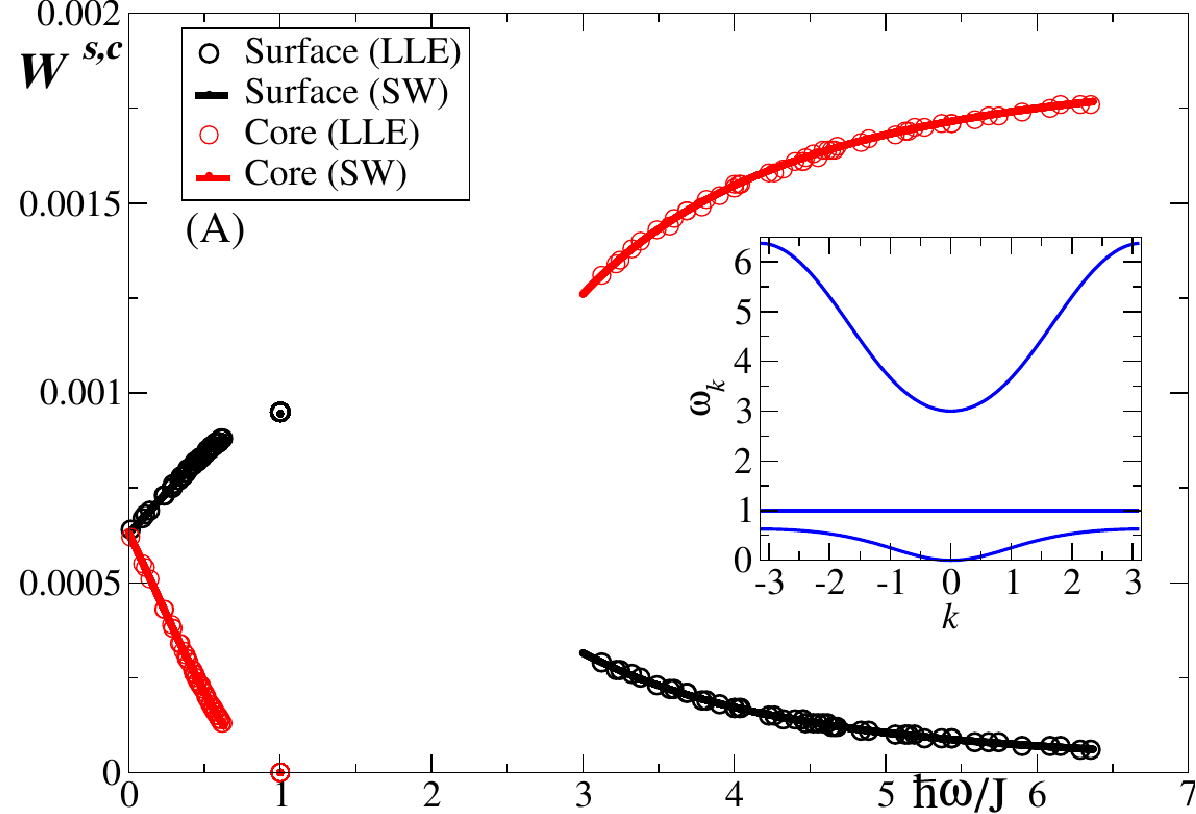}\includegraphics[height=4.6cm]{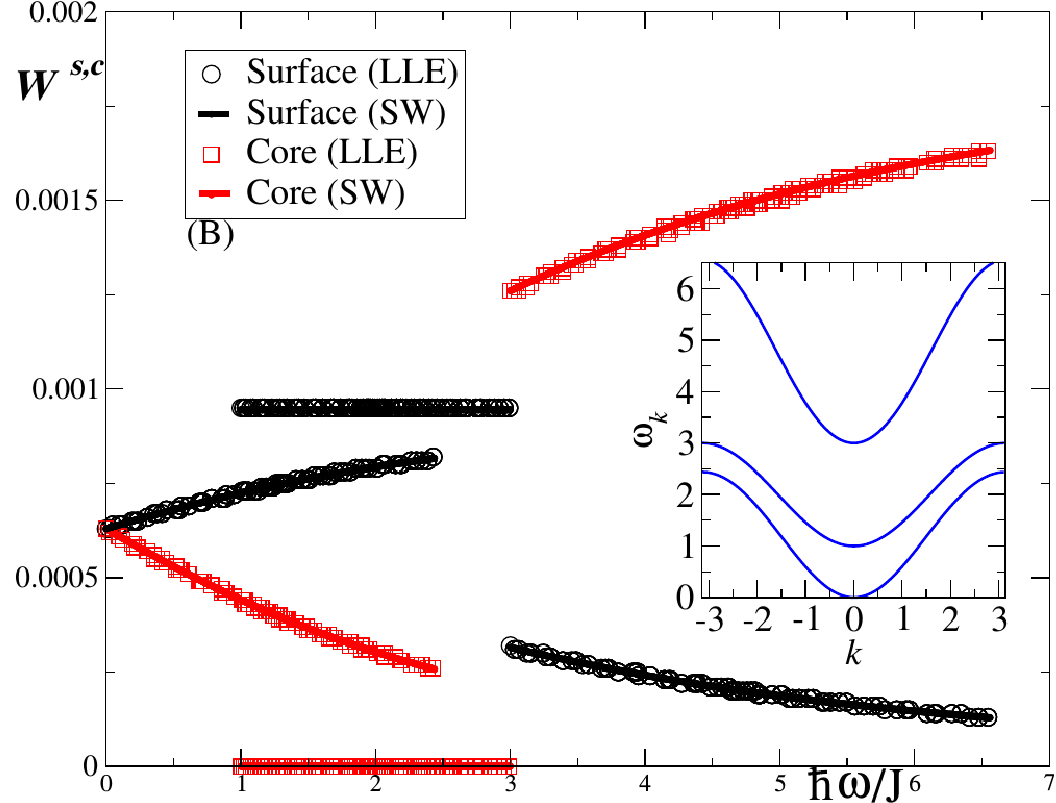} 
\par\end{centering}

\noindent \centering{}\protect\caption{Surface and core spectral weights against the magnon energy for $j_{c}=j_{cs}=1$
and with $j_{s}=0\ (A),\ j_{s}=0.5\ (B)$. \label{fig:energy_spectrum_qxqy-1} }
\end{figure*}

Then, we rewrite the Hamiltonian (\ref{eq:hamiltonian-heisenberg})
in terms of the real-space magnon operators $a_{i,l}$ and $a_{i,l}^{\dagger}$.
The resulting expression can be partially diagonalized after a Fourier
transformation with respect to the $\left(x,y\right)$ directions
\begin{equation}
\begin{array}{lll}
\frac{\mathcal{H}}{S} & = & {\displaystyle \sum_{q_{x},q_{y}}}\left(a_{q,1}^{\dagger}\ a_{q,2}^{\dagger}\ a_{q,3}^{\dagger}\right)\mathcal{J}\left(\bm{q}\right)\cdot\left(\begin{array}{c}
a_{q,1}\\
a_{q,2}\\
a_{q,3}
\end{array}\right)+{\rm C^{te}},\end{array}\label{eq:hamiltonian-magnon}
\end{equation}
where $\mathcal{J}\left(\bm{q}\right)$ is the coupling matrix 
\begin{equation}
\mathcal{J}\left(\bm{q}\right)=\left(\begin{array}{ccccc}
\mathcal{J}_{11} &  & -J_{cs} &  & 0\\
-J_{cs} &  & \mathcal{J}_{22} &  & -J_{cs}\\
0 &  & -J_{cs} &  & \mathcal{J}_{33}
\end{array}\right),\label{eq:coupling_matrix}
\end{equation}
with $\mathcal{J}_{11}=\mathcal{J}_{33}=2J_{s}\left(1-\gamma_{\bm{q}}\right)+J_{cs}$
and $\mathcal{J}_{22}=2J_{c}\left(1-\gamma_{\bm{q}}\right)+2J_{cs}$,
and $\gamma_{\bm{q}}\equiv\frac{1}{2}\left(\cos q_{x}+\cos q_{y}\right)$.
We use $J_{c}$ as our energy scale and define the reduced couplings
$j_{cs}\equiv J_{cs}/J_{c}$ and $j_{s}\equiv J_{s}/J_{c}$. The three
dispersions are then given by 
\begin{equation}
\begin{array}{lll}
\omega_{\pm}\left(j_{cs},j_{s},\bm{q}\right) & = & \frac{3}{2}j_{cs}+\left(1+j_{s}\right)\left(1-\gamma_{\bm{q}}\right)\\
\\
 &  & \pm\frac{1}{2}\left\{ \left[3j_{cs}+2\left(1+j_{s}\right)\left(1-\gamma_{\bm{q}}\right)\right]^{2}\right.\\
 &  & \left.\qquad-8\left(1-\gamma_{\bm{q}}\right)\left[j_{cs}\left(1+j_{s}\right)+2j_{s}\left(1-\gamma_{\bm{q}}\right)\right]\right\} ^{1/2}\\
\\
\omega_{0}\left(j_{cs},j_{s},\bm{q}\right) & = & j_{cs}+2j_{s}\left(1-\gamma_{\bm{q}}\right).
\end{array}\label{eq:energy_dispersion}
\end{equation}

The spectral weights are then obtained as the squares of the projections
of the eigenvectors onto the canonical basis $e_{i}^{\alpha}=\delta_{i}^{\alpha},i=1,2,3;\alpha=x,y,z$.
These weights depend on the physical parameters such as the exchange
couplings and anisotropy constants. Upon summing over the wave vectors
$\bm{q}$ within the first Brillouin zone, one can plot the spectral
weights as functions of $\omega\left(\bm{q}\right)$.

In Fig. \ref{fig:energy_spectrum_qxqy-1} we present the spectral
weight of the surface and core spins as a function of the magnon energy
for the three energy bands, along the path $q_{x}=q_{y}$, corresponding
to the three dispersions (\ref{eq:energy_dispersion}). The circles
and squares are the results for a finite cluster ($N_{x}=N_{y}=23$)
dealt with using the numerical method of Subsection \ref{sub:LinLLE},
with periodic boundary conditions in the $x$ and $y$ directions.
The full lines are the results obtained within the spin-wave approach
presented above. The results in Fig. \ref{fig:energy_spectrum_qxqy-1}
exhibit a very good agreement between the numerical and analytical
approaches for all values of the exchange parameters.

In the spin-wave calculation we consider blocks of three spins, belonging
to layers 1, 2, 3. These blocks are coupled to one another by lateral
(in-plane) couplings. The spin-wave dispersion, as shown in the inset
of Fig. \ref{fig:energy_spectrum_qxqy-1}, has three branches: the
lowest branch corresponds to the ferromagnetic magnon excitations
with the 3 spins precessing in phase. By computing the spectral weight
associated with this branch, one finds that the surface contribution
dominates (apart from the uniform mode at $\bm{k}=0$) because the
corresponding modes require less energy to be excited. In contrast,
the high-energy branch corresponds to the situation where the end
spins (layers) precess with opposite phases. The spectral weight is
then dominated by the core owing to a higher spin stiffness. For the
particular case of $j_{s}=0$, the magnon dispersion exhibits a non-dispersive
branch at $\omega_{k}=1$ {[}see inset of Fig. \ref{fig:energy_spectrum_qxqy-1}
(left){]}. This intermediate branch follows from the fact that the
bottom and top layer spins are not coupled within their respective
planes. Therefore, creating an excitation within the top or bottom
layer is costless, leading to a mode with constant energy in $k$-space.
Obviously, this branch corresponds to excitations that are localized
at the surface. This can be seen by examining the spectral weight
for which the core contribution vanishes.

As the surface exchange coupling increases (\emph{i.e.} $j_{s}>0$)
more dispersion is observed and the branches start to merge for some
magnon energies. Hence, the spectral weight changes both qualitatively
and quantitatively: the gaps close and the surface and core contributions
become more and more entangled.

\textcolor{black}{The calculation of the absorbed power for this system
yields one absorption peak for the uniform mode corresponding to the
lower energy band in Fig.\ref{fig:energy_spectrum_qxqy-1}. The eigenfunctions
for the three energy bands at $\bm{k=0}$ are given by}

\begin{equation}
\left\{ \begin{array}{lll}
\Psi_{1} & = & \frac{1}{\sqrt{3}}\left(\phi_{S1}+\phi_{C}+\phi_{S2}\right),\\
\Psi_{2} & = & \frac{1}{\sqrt{2}}\left(\phi_{S1}-\phi_{S2}\right),\\
\Psi_{3} & = & \frac{1}{\sqrt{6}}\left(\phi_{S1}-2\phi_{C}+\phi_{S2}\right).
\end{array}\right.
\end{equation}

\textcolor{black}{Here $\phi_{S1,2}$ corresponds to the surface spins
and $\phi_{C}$ to the core spin. The coefficients $C_{k,\ell}$ of
these vectors do not vanish (and are all equal) for the vector $\Psi_{1}$
that corresponds to the uniform mode. In order to obtain more absorption
peaks in the absorbed power we can introduce a core anisotropy $k_{c}$
but no surface anisotropy. In this case the eigenfunctions corresponding
to $\bm{k=0}$ are 
\begin{equation}
\left\{ \begin{array}{lll}
\Psi_{1} & = & \frac{2}{N_{1}}\left[\phi_{S1}-\left(1+k_{c}-\sqrt{9+2k_{c}+k_{c}^{2}}\right)\phi_{C}+\phi_{S2}\right],\\
\Psi_{2} & = & \frac{1}{\sqrt{2}}\left(\phi_{S1}-\phi_{S2}\right),\\
\Psi_{3} & = & \frac{2}{N_{3}}\left[\phi_{S1}-\left(1+k_{c}+\sqrt{9+2k_{c}+k_{c}^{2}}\right)\phi_{C}+\phi_{S2}\right],
\end{array}\right.
\end{equation}
where $N_{1}$ and $N_{3}$ are normalization factors of the wave-vectors
$\Psi_{1}$ and $\Psi_{3}$ respectively. We can see that the modes
corresponding to $\Psi_{1}$ and $\Psi_{3}$ can contribute to the
absorbed power. }

\bibliographystyle{iopart-num}
\bibliography{new-biblio}

\end{document}